\documentclass[useAMS,usenatbib,usegraphicx]{mn2e}
\usepackage{subfigure}      

\title[]{The role of the initial surface density profiles of the disc on giant planet formation: comparing with observations}
\author[Y. Miguel, O. M. Guilera and A. Brunini]{Y. Miguel$^{1,2}$\thanks{E-mail: ymiguel@fcaglp.unlp.edu.ar}, O. M. Guilera$^{1,2}$ and A. Brunini$^{1,2}$\thanks{Member of the Carrera del Investigador Cient\'\i fico. Consejo Nacional de Investigaciones Cient\'\i ficas y T\'ecnicas (CONICET).E-mail: abrunini@fcaglp.unlp.edu.ar}\\
$^1$Facultad de Ciencias Astron\'omicas y Geof\'\i sicas. Universidad
Nacional de La Plata. Paseo del Bosque s/n, La Plata (1900), Argentina.\\
$^2$Instituto de Astrof\'\i sica de La Plata (CCT La Plata-CONICET, UNLP), Paseo del Bosque s/n, La Plata (1900), Argentina}

\begin{document}  

\pagerange{\pageref{firstpage}--\pageref{lastpage}}

\label{firstpage}

\maketitle
        
\begin{abstract}

In order to explain the main characteristics of the observed population of extrasolar planets and the giant planets in the Solar System, we need to get a clear understanding of which are the initial conditions that allowed their formation. To this end we develop a semi-analytical model for computing planetary systems formation based on the core instability model for the gas accretion of the embryos and the oligarchic growth regime for the accretion of the solid cores. With this model we explore not only different initial discs profiles motivated by similarity solutions for viscous accretion discs, but we also consider different initial conditions to generate a variety of planetary systems assuming a large range of discs masses and sizes according to the last  results in protoplanetary discs observations. We form a large population of planetary systems in order to explore the effects in the formation of assuming different discs and also the effects of type I and II regimes of planetary migration, which were found to play fundamental role in reproducing the distribution of observed exoplanets. Our results show that the observed population of exoplanets and the giant planets in the Solar System are well represented when considering a surface density profile with a power law in the inner part characterised by an exponent of $-1$, which represents a softer profile when compared with the case most similar to the MMSN model case. 

\end{abstract}

\begin{keywords} 
Planets and satellites: formation\ - Solar System: formation\
\end{keywords}

\section{Introduction}

So far 464 planets have been found orbiting stars in the solar neighbourhood. This population of planets, though strongly biased to those planets that are easier to be detected with radial velocities, present tendencies that the theoretical models of planetary formation must reproduce. With the end of explaining the main characteristics of the observed population of extrasolar planets, including the planets in the Solar System, we need to get a clear understanding of which are the initial conditions that allow their formation.

The fact that the planetary systems formed from a disc-like nebula was recognized centuries ago, but even today our understanding of this processes and initial conditions remain uncertain. With this question in mind, the main objective of this work is to determine which is the nebula model that allows us to form planetary systems whose planets reproduce the observational sample of extrasolar planets.

In recent years there have been several semi analytical models of planetary formation \citep{b16,b8,b21} and planetary systems formation \citep{b9}, which intend to reproduce the observational sample of exoplanets and explain with a simple model the main characteristics of this distribution in order to get a better understanding of the process of planetary formation. For simplicity the surface density model assumed in these models as basic building blocks for the planetary formation process are based on conventional power-law models, as the standard minimum mass solar nebula (MMSN) model of \citet{b1}, which is an approximation to the gas-dominated nebula, taking into account the primitive composition of the solar disc. This model, although efficient on its simplicity, suffers from multiple disadvantages.

The minimum nebula \citep{b36,b37,b14,b38,b1}, was constructed by assuming that there should be added to the masses of the present planetary bodies enough icy materials and hydrogen and helium to achieve solar composition and then this mass should be smeared over the surrounding place. So the model was based on the strong implicit assumption that planets were formed at their present locations, they accreted all of the solids in their vicinity and the accretion was perfect, which means that the formation of one giant planet did not have consequences on the formation of the others. Despite this model is a reasonable good approximation in the intermediate region of the solar nebula, posses too much surface density in the inner region, under predicts it in the outer nebula, and the total mass of the disc is infinite in this formulation, so inner and outer boundaries must be specified. For all these reasons and despite its widespread use, it is clear that this model must be updated or changed.

With the spirit of retaining a simple model for the surface density in the nebula and get a more realistic model consistent with protoplanetary discs observations, we consider a nebula with a surface density profile motivated by similarity solutions for viscous accretion discs as shown in \citet{b4} and in \citet{b5}. This simple formula, which is characterised by a power-law in the inner part of the disc and an exponential decay in the outer part, is a reasonable alternative for use in protoplanetary discs because it is simple, it matches the protoplanetary discs observations \citep{b2,b3} and unlike power-law approximations, these surface density model predicts a sharp outer edge to the nebula without introducing an arbitrarily cut to the disc. There are several free parameters in the nebula model that we have to adjust with the aim of explaining the distribution of observed exoplanets and the giant planets in our Solar System, such as the exponent of the power-law in the inner part of the disc, whose different values will bring a significant impact on the formation of giant planets as we discuss in this work.   

So with the aim of finding the planetary system formation model that best fits the observations (including our Solar System), we have developed a semi analytical model where the main difference with our previous model \citep{b9} is the initial nebula profile. The model, which is shown in section \ref{modelo}, is based on the nucleated instability model, where the solid cores grow in the oligarchic growth regime. We also include the effects of having embryos in a gaseous disc, considering type I and II regimes of planetary migration, which were found to be fundamental in reproducing the distribution of observed exoplanets. The results are shown in section \ref{resultados} where we also show the comparison with the observations and the summary and conclusions are shown in section \ref{conclusion}.       

\section{Model and basic equations} \label{modelo}

In this section we show the model considered, which is based on a model previously developed by \citet{b8,b9,b10}. As the main goal of this work is to show which are the consequences of considering different density profiles and which one suits better the observations and the Solar System, we will explain the protoplanetary nebula model in detail.

\subsection{Protoplanetary Nebula Structure}\label{nebulosa}

The structure of the protoplanetary nebula is usually assumed to follow a power-law distribution in the form,
 
\begin{equation}
\Sigma(a)=\Sigma_0\bigg(\frac{a}{a_0}\bigg)^{-p}
\end{equation}
where $\Sigma_0$ is the surface density at the arbitrary radius $a_0$. In this formulation, inner and outer boundaries of the disc must be arbitrarily specified  otherwise the total mass of the disc would be infinite. This is one of the limitations suffered by this formulation, that was initially motivated by models as the minimum mass solar nebula model (MMSN) developed by \citet{b1}.  

In order to consider a more appropriate disc structure model, we are based on the works of \citet{b2} and \citet{b3}, who adopted a density profile characterised by a power-law in the inner part of the disc and an exponential decay in the outer parts of the disc and is based on the similarity solutions of the surface density of a thin Keplerian disc subject to the gravity of a point mass ($M_{\star}$) central star \citep{b4,b5}. In this formulation the gas surface density is given by,

\begin{equation}\label{dengas}
\Sigma_g(a)=\Sigma_g^0 \bigg(\frac{a}{a_c}\bigg)^{-\gamma} e^{-\big(\frac{a}{a_c}\big)^{2-\gamma}}
\end{equation}   
where $a_c$ is a characteristic radius beyond which the density drops sharply, $\gamma$ is the exponent that defines the density profile and $\Sigma_g^0$ is a parameter that is determined from the total mass of the disc which is given by, 

\begin{equation}
M_d=2\pi\int_0^{\infty}\Sigma_g(a)\,a\, da
\end{equation}
solving the integral we get the expression for $\Sigma_g^0$, 

\begin{equation}
\Sigma_g^0= \frac{(2-\gamma)M_d}{2\pi a_c^2}, \qquad \mbox{with}~\gamma < 2
\end{equation}

We adopted a temperature profile for a disc optically thin, given by

\begin{equation}
 T=280\bigg(\frac{a}{1~au}\bigg)^{-1/2}\bigg(\frac{L_{\star}}{L_{\odot}}\bigg)^{1/4} \, K
\end{equation}
with $L_{\odot}$ and $L_{\star}$ the Sun and Stellar luminosity. 
The volumetric density of gas is 

\begin{equation}
\rho=\rho_0 e^{-\big(\frac{Z}{h}\big)}
\end{equation}
where $h$ is the disc scale height ($h \simeq 0.05(a/1au)^{1/4}a$) and $\rho_0$ is the volumetric density of gas in the midplane of the disc given by

\begin{displaymath}
\rho_0=8.33 \times 10^{-13}\bigg(\frac{M_{\star}}{M_{\odot}}\bigg)^{1/2}\frac{\Sigma_g^0}{a_c^{5/4}}\bigg(\frac{a}{a_c}\bigg)^{-\gamma-5/4}
\end{displaymath}
\begin{equation}
e^{-\big(\frac{a}{a_c}\big)^{2-\gamma}}\bigg(\frac{L_{\odot}}{L_{\star}}\bigg)^{1/8}\, g\, cm^{-3}
\end{equation}   

We will assume that the solids surface density has an expression similar to equation \ref{dengas}, 

\begin{equation}
\Sigma_s(a)=\Sigma_s^0 \eta_{ice}\bigg(\frac{a}{a_c}\bigg)^{-\gamma} e^{-\big(\frac{a}{a_c}\big)^{2-\gamma}}
\end{equation}   
where $\eta_{ice}$ is a function introduced in order to represent the change in the solids beyond the radius where the water condenses,

\begin{equation}
\eta_{ice} = \left\{ \begin{array}{ll}
1 & \textrm{if $a>a_{ice}$}\\
\frac{1}{4} & \textrm{if $a\leq a_{ice}$}
\end{array} \right.
\end{equation} 
where the snow line is located at $a_{ice}=2.7\big(\frac{M_{\star}}{M_{\odot}}\big)^2$ $au$ from a central star of mass $M_{\star}$ and if $z_0$ is the primordial abundance of heavy elements in the Sun and we assume the same for the disc, then  

\begin{equation}
\bigg(\frac{\Sigma_s^0}{\Sigma_g^0}\bigg)_{\odot}=z_0
\end{equation}
following the result of \citet{b6} we assumed that $z_0=0.0149$ and for a star of metallicity $[Fe/H]$ then
\begin{equation}
\bigg(\frac{\Sigma_s^0}{\Sigma_g^0}\bigg)_{\star}= \bigg(\frac{\Sigma_s^0}{\Sigma_g^0}\bigg)_{\odot}10^{[Fe/H]}= z_0 10^{[Fe/H]}
\end{equation}
The metallicities are taken random considering that they follow a log-normal distribution fitted from the results of the CORALIE sample, as is shown in \citet{b21}.

\begin{figure}
  \begin{center}
    \includegraphics[angle=270,width=.5\textwidth]{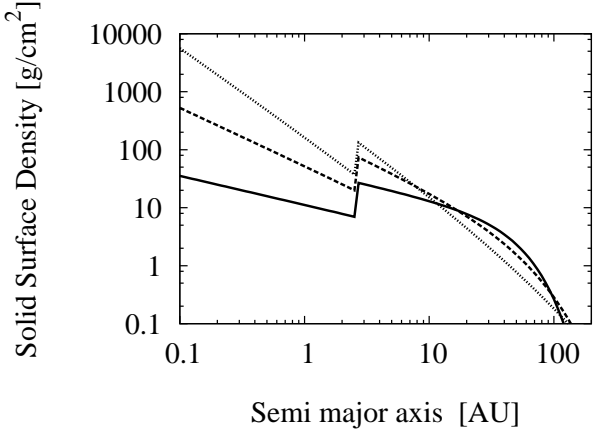}
  \end{center}
  \caption{Solid surface density for three discs with the same total mass, characteristic radius and metellicity but different values for $\gamma$. The dotted line represents $\gamma=1.5$, the dashed line is for $\gamma=1$ and the solid line represents the density for $\gamma=0.5$.}
  \label{densol}
\end{figure}

In order to illustrate we show in Figure \ref{densol} the solids surface density as a function of semi major axis for three discs with $a_c = 50~au$, total mass of $0.05~M_{\odot}$, solar metallicity  and different values for $\gamma$. The dotted line represents the solid surface density for $\gamma=1.5$,
the dashed line shows the curve for the case when $\gamma=1$ and finally the solid line is the solid surface density for $\gamma=0.5$. We note that as the value of the exponent $\gamma$ decreases, the amount of solids in the inner part of the disc is reduced, while it increases in the outer part, fact that brings deep consequences on the formation of giant planets, as we will see in the results. The case most similar to the MMSN model is when $\gamma=1.5$ where we note an excess of solids in the inner part of the disc, which drops sharply beyond the snow line. This case is less favorable for the formation of several giant planets in the same planetary system, since the mass is concentrated and as a consequence the formation of a single giant planet in the snow line is allowed leaving no residual material for others after its formation. Recently \citet{b7} have shown that when employing power-law discs $\Sigma \propto a^{-p}$, the value $p= 1.5$ leads to a quickly formation of Jupiter that could inhibit the formation of Saturn when the formation of both is considered simultaneously. They also showed that smoother surface density profiles (p=1, 0.5) favoured the simultaneous formation of Jupiter and Saturn.

The disc is extended between $a_{int}$ and $a_{ext}$, where the inner radius is calculated accorded to an expression given by \citet{b11} who found it through observations in protoplanetary discs,

\begin{equation}\label{innerradius}
a_{in}=0.0688 \bigg(\frac{1500^{\circ}K}{T_{sub}}\bigg)^2\bigg(\frac{L_{\star}}{L_{\odot}}\bigg)^{\frac{1}{2}} au
\end{equation}
with $T_{sub}$ the dust sublimation temperature taken as $1500^{\circ}K$ and $L_{\star}$ and $L_{\odot}$ are the stellar and Sun luminosity respectively. The inner radius takes values around $0.01~au$. The outer radius is the radius that contains $95\%$ of the disc mass (the total disc mass is aproximately the gaseous mass of the disc),

\begin{equation}
0.95M_d\simeq 2\pi \int_{0}^{a_{ext}}\Sigma_g(a)a~da 
\end{equation}
where we assume that the total disc mass is aproximately the gaseous mass of the disc. Then,
\begin{equation}
a_{ext}=3^{\frac{1}{2-\gamma}}a_c
\end{equation}
which for a disc characterised by $\gamma=1$ ranges between $\sim 90$ and $\sim 300~au$.

The total mass of the discs and their characteristic radius are taken random considering that they also follow a log-Gaussian distribution that we fitted following the results of \citet{b2} and \citet{b3}. As the masses of the generated discs could be large enough to undergo in a gravitational instability we check that our discs are stable. The gravitational stability of a Keplerian accretion disc with sound speed $c_s$ is measured by the Toomre Q-parameter \citep{b39}, which is defined by,

\begin{equation}\label{q1}
Q=\frac{c_s \, \Omega_K}{\pi\, G\, \Sigma_g}
\end{equation}
In the case of our disc model,

\begin{equation}\label{q}
Q \simeq 1.24 \times 10^5 \bigg(\frac{a}{1au}\bigg)^{\gamma-\frac{7}{4}}\bigg(\frac{a_c}{1au}\bigg)^{-\gamma}\bigg(\frac{M_{\star}}{M_{\odot}}\bigg)\frac{e^{(\frac{a}{a_c})^{2-\gamma}}}{\Sigma_g^0}
\end{equation}
where a value of $Q \leq 1$ represents an unstable disc. 

As we note in the previous equation, the parameter of instability depends on the semimajor axis, therefore it changes through the disc. The minimun of equation \ref{q} is found when $a = \big(\frac{\frac{7}{4}-\gamma}{2-\gamma}\big)^{\frac{1}{2-\gamma}}a_c$, so we chose $Q_{a_{min}}$ (the value of Q when $a = a_{min}$)  as a representative value for the disc. Those discs with $Q_{a_{min}} > 1$ will be stable all over the disc. We also notice that the gravitational stability parameter depends on $\gamma$, hence different values of $\gamma$ lead to different values of the Toomre parameter and therefore two discs with equal mass and characteristic radius could be stable or unstable, depending on the density profile. 

\begin{figure}
  \begin{center}
    \subfigure[]{\label{primerag}\includegraphics[angle=270,width=.48\textwidth]{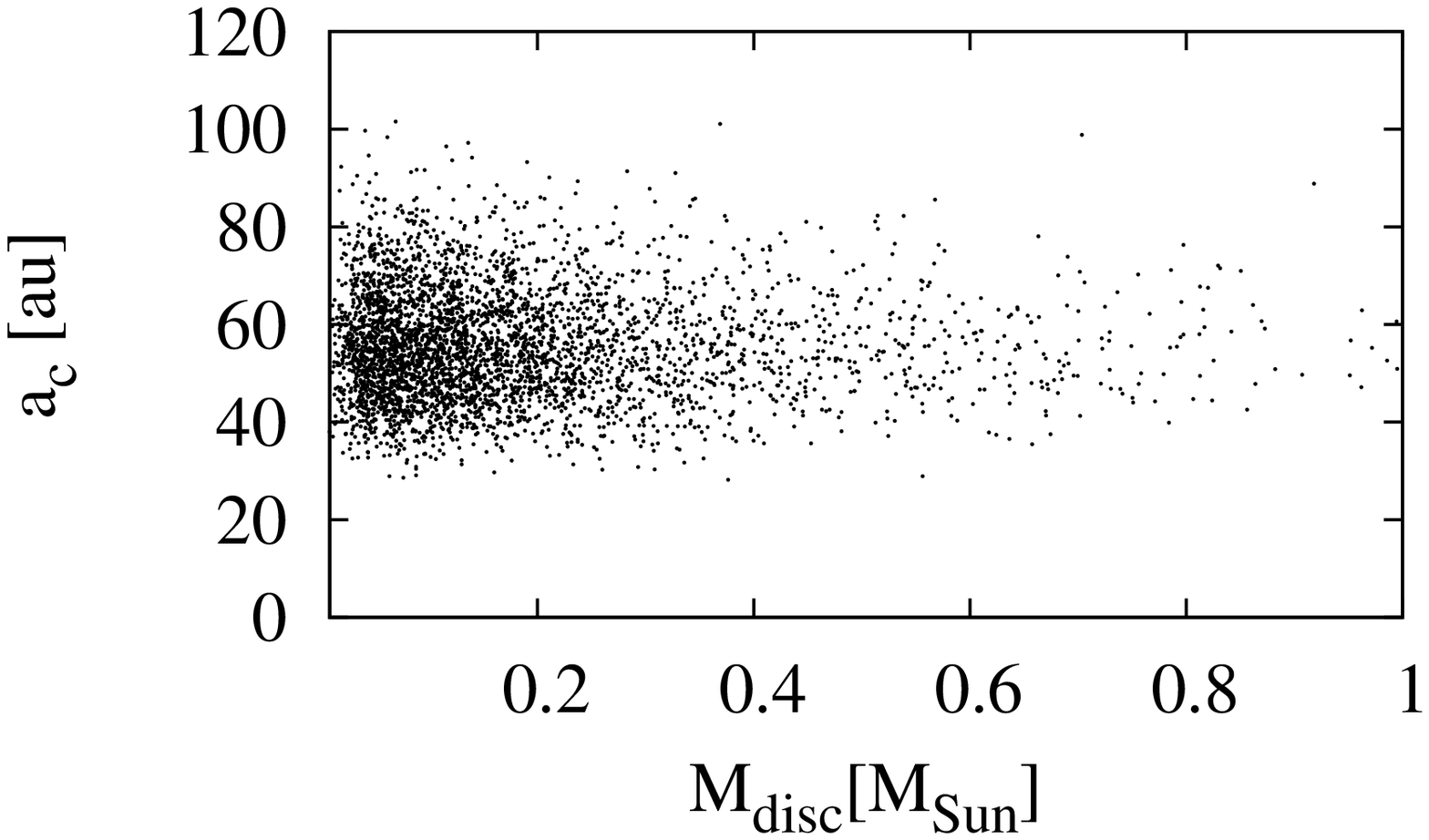}}
    \subfigure[]{\label{segundag}\includegraphics[angle=270,width=.48\textwidth]{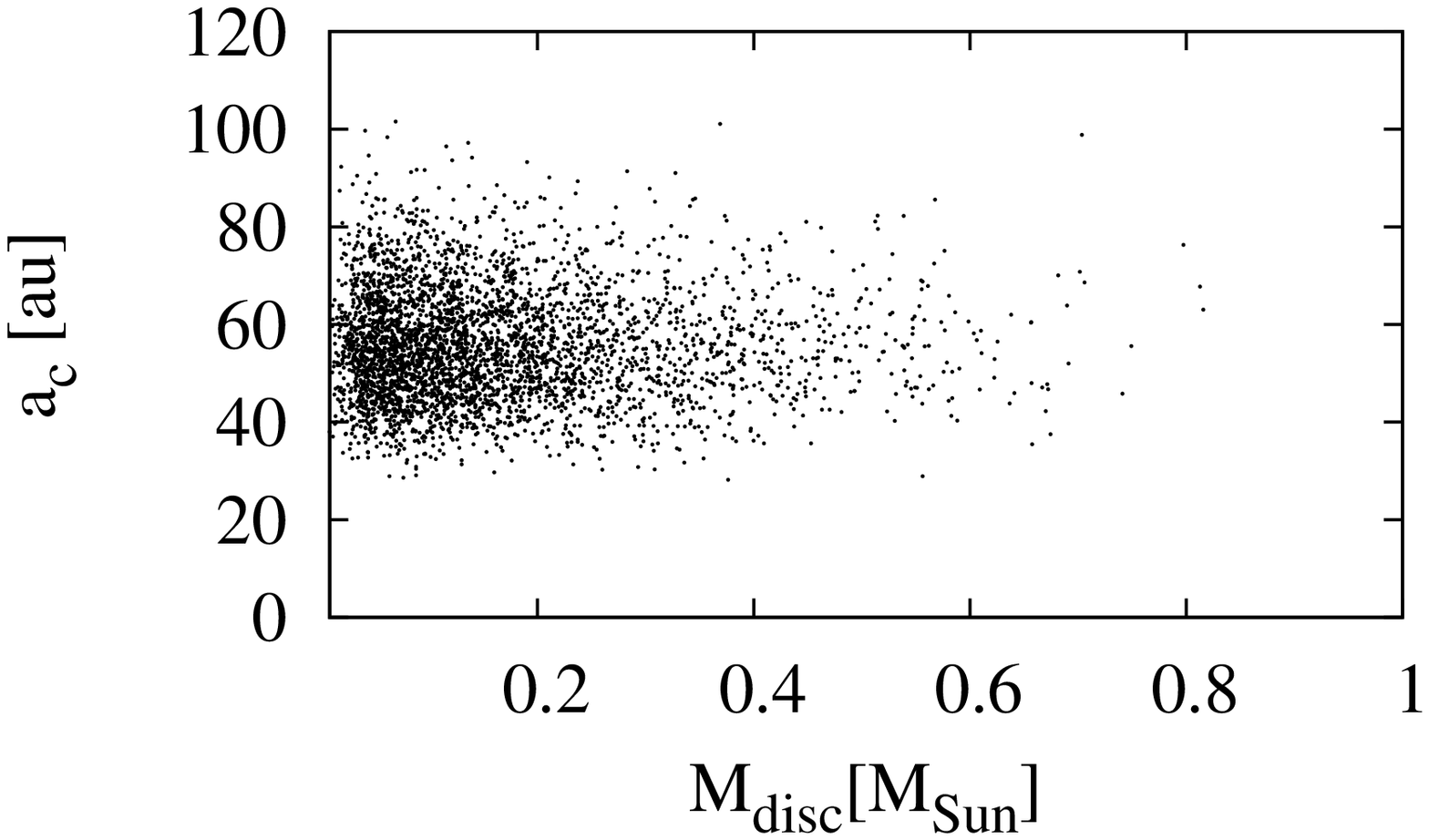}}
    \subfigure[]{\label{tercerag}\includegraphics[angle=270,width=.48\textwidth]{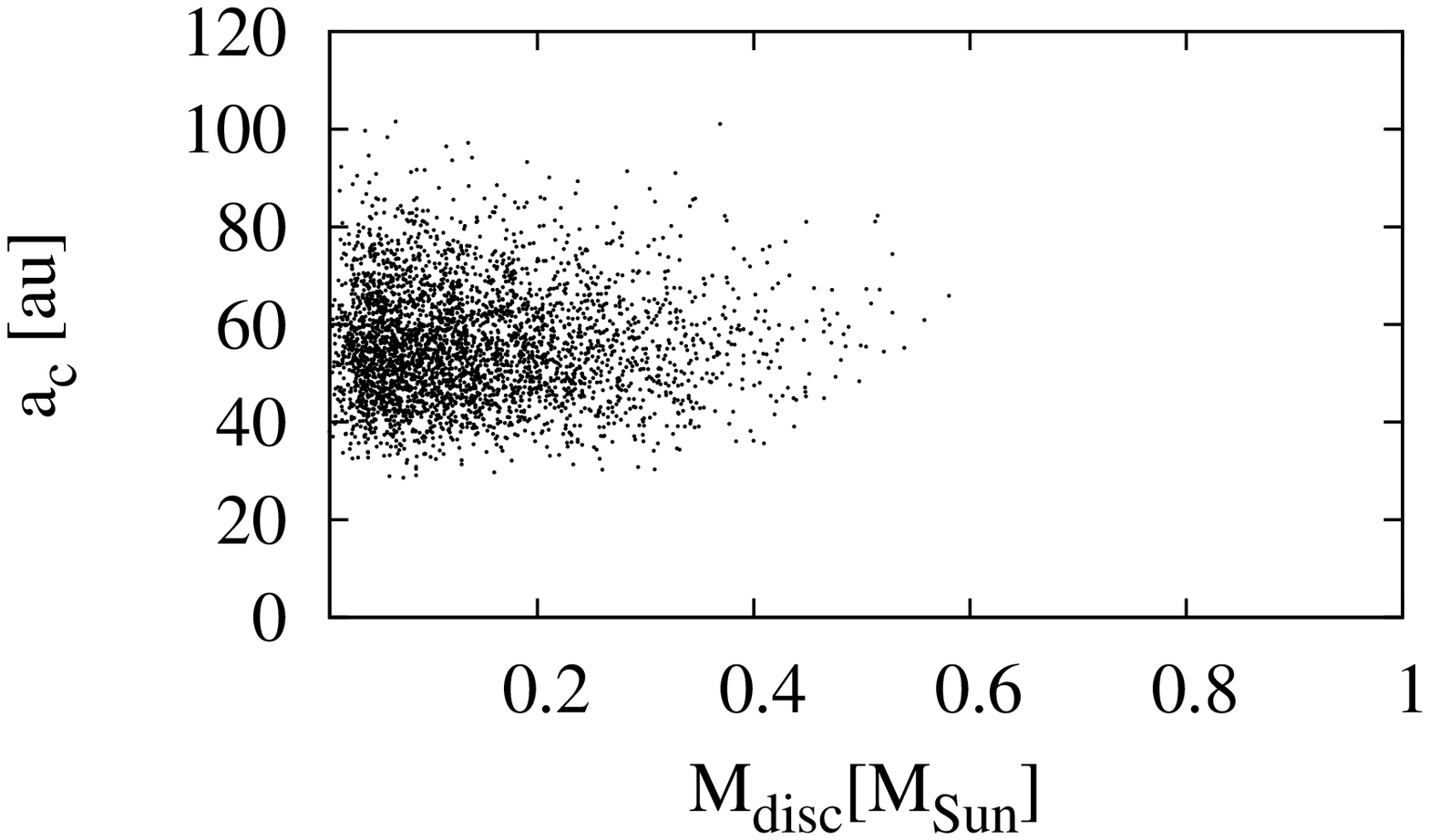}}
  \end{center}
  \caption{Mass and characteristic radius of all the discs generated with $Q_{amin}>1$. the first figure shows the results when we consider $\gamma=1.5$, figure \ref{segundag} shows the stable discs generated when $\gamma= 1$ and finally stable discs with $\gamma=0.5$ are shown in figure \ref{tercerag}   }
  \label{estables}
\end{figure}

This is shown in figures \ref{estables} where we show the mass vs. the $a_c$ of all the stables discs generated, and the different figures show the results for three different values of the parameter $\gamma$. Figure \ref{primerag} show the stable discs generated with a density profile corresponding to $\gamma=1.5$, \ref{segundag} are the discs with $\gamma=1$ and \ref{tercerag} when $\gamma=0.5$. 
We note that for larger values of $\gamma$ we need more mass to obtain a
gravitationally unstable disc. This suggest that, as is seen in equation
\ref{q1}, the global nature of the instability depends both on the mass of the
disc and on how this mass is distributed.

Finally, in figures \ref{primerag} and \ref{segundag}, we note the presence of
extremely massive discs. These discs should not be considered as keplerian and
equation \ref{q1} does not apply to them. In order to avoid these very massive
discs, we check that the disc mass is less than 20\% of the mass of the central
star \citep{b5}. Thus we consider discs with masses up to $0.28~M_{\odot}$.
 
In order to represent in a simplified way the depletion of the gaseous disc, we assume that this mechanism can be modeled considering an exponential decay for the mass of the gaseous disc, which occurs on time scales between $10^6$ and $10^7$ years in agreement to observation of circumstellar discs \citep{b41,b42}.

\subsection{Growth of the Protoplanetary Embryos}

Once we have the protoplanetary nebula defined, we locate the initial cores. The first one is located at $a_{int}$ and the others are located at 10 Hill radii away from each other until it reaches the disc outer radius $a_{ext}$. Each initial embryo has the minimum mass necessary for starting the oligarchic growth regime \citep{b12,b13}, which is the growth stage where our embryos grow.  

\begin{equation}\label{Masa-inicial}
M_{oli}\simeq \frac{1.6 a^{\frac{6}{5}}10^{\frac{3}{5}}m^{\frac{3}{5}}\Sigma_s^{\frac{3}{5}}}{M_{\star}^{\frac{1}{5}}}
\end{equation}
with m the effective planetesimal mass.  

\subsubsection{Solid Accretion into the cores}\label{solidos}

The cores grow due to the accretion of solids, gas and also due to the collision with other embryos. The rate at which a core accumulates solids in the oligarchic growth regime was found by \citet{b14} and has the form,  

\begin{equation}
\frac{dM_{s}}{dt}=10.53 \Sigma_s \, \Omega \, R_p^2\bigg(1+\frac{2GM_t}{R_p\sigma}\bigg)
\end{equation}
where $\Omega$ is the Kepler frequency, $R_p$ and $M_t$ are the planet's radius and total mass (solid + gas) and $\sigma$ is the velocity dispersion which depends on the eccentricity of the planetesimals in the disc. We assume that the rms eccentricity of the planetesimals in the disc are damped and have reached an equilibrium value which is, 

\begin{equation}\label{em}
e_m^{eq}=\frac{1.7\, m^{1/15} \, M_t^{1/3} \, \rho_m^{2/15}}{10^{1/5}\, \rho_{0}^{1/5}\, M_{\star}^{1/3}\, a^{1/5}}
\end{equation}
as is shown in \citet{b15}, where $\rho_m$ is the planetesimal bulk density. With this expression the solid accretion rate including the evolution of the planetesimal rms $e$ and $i$ has the form, 

\begin{equation}\label{core-accretion}
\frac{dM_s}{dt}\simeq \frac{3.9 10^{\frac{2}{5}}G^{\frac{1}{2}}M_{\star}^{\frac{1}{6}}\rho_{0}^{\frac{2}{5}}\Sigma_s}{\rho_m^{\frac{4}{15}}\rho_M^{\frac{1}{3}}a^{\frac{1}{10}}m^{\frac{2}{15}}}M_t^\frac{2}{3}
\end{equation}
where $\rho_M$ is the embryo bulk density, which is equal to the planetesimals density, $\rho_M=\rho_m=\rho$.

The solids accretion ends when $\Sigma_s$ is zero in their feeding zones. The solids surface density diminishes due to a combination of two factors:
\begin{itemize}
\item{} the cores ate the solids in their feeding zones,
\item{} the ejection of planetesimals diminished the solids surface density in the region \citep{b15,b16}.
\end{itemize}

We also consider that when a core is able to retain gas, the drag effect caused by the gaseous envelope on the planetesimals increases the collision cross section of the embryo. Following \citet{b45}, we assume that the enhanced collision radius ($R_{collision}$) of the embryo is given by,

\begin{equation}
\bigg(\frac{R_{collision}}{R}\bigg)^4=\frac{0.000344\mu^4cP}{\kappa r_m \Sigma_d}\bigg(\frac{M_t}{M_{\oplus}}\bigg)^2\bigg(\frac{24e_{eq}^2}{24+5e_{eq}^2}\bigg)
\end{equation}
where $c$ is the velocity of light, $P$ is the orbital period of the protoplanet, $\kappa$ is the opacity of the atmosphere which is considered as $\simeq 4 cm^2 g^{-1}$, $\mu\simeq 2.8$ is the mean molecular weight of the atmosphere (assumed to have a solar composition)  and the embryos equilibrium eccentricity is considered as $\approx 2$ in this expression. 

As was said in the beginning of this section, collisions represent an important evolutionary process which plays a significant role in determining the final mass and spin state of the planets. The model and consequences of embryo-embryo collisions was explained in a previous work \citep{b10}, here we say that the result of the collisions are the perfect accretion of the embryos involved.

\subsubsection{Gas accretion onto the Cores}\label{gas}

The cores have  an associate  envelope if  the molecular  velocity is smaller than the escape one. According to \citet{b46}, once a protoplanet becomes greater than  $\sim$ the Moon's mass,  the core  attracts the neighboring  gas and  an envelope forms surrounding it. In the early stages of giant planets formation, the gravity is  balanced by the pressure gradient  which is maintained by  the potential  energy released  by incoming  planetesimals \citep{b47}, so the stability of this  envelope depends on the mass of the protoplanet and on the solids accretion rate. 
When the mass of the embryo becomes greater than a critical value, or there are no more solids available on its feeding zone, the envelope can no longer be in hydrostatic equilibrium and begins to collapse \citep{b47,b48}, as a result, the gas accretion process began. The core' s critical mass \citep{b17,b40} is given by, 

\begin{equation} \label{mcritica}
M_{crit}\sim 10 \bigg(\frac{\dot{M_c}}{10^{-6}M_{\oplus}yr^{-1}}\bigg)^{\frac{1}{4}}
\end{equation} 

\citet{b18} and \citet{b19} studied the formation and evolution of a protoplanet in situ, in the  frame of the nucleated instability model. They improved a numerical code based on a Henyey technique \citep{b20}, and consider that the oligarchic growth regime is the stage for the accretion of the solid cores and considered the gas drag effect acting on planetesimals inside the planet atmosphere. We used the results of \citet{b18} introducing an analytic approximation to their numerical results as is explained in \citet{b8}, and update here the gas accretion rate according to their last results \citep{b19}. Then the gas accretion process occurs on a rate,

\begin{equation} \label{acregas1}
\frac{dM_g}{dt}=\frac{M_t}{\tau_g}
\end{equation}
where $M_g$ is the mass of the surrounding envelope and $\tau_g$ is its characteristic growth time, 

\begin{equation} \label{acregas2}
\tau_g=8.35 \times 10^{10} \bigg(\frac{M_t}{M_{\oplus}}\bigg)^{-4.89}yrs
\end{equation}

We assume the next limits for ending the gas accretion process:
\begin{itemize}
\item the growth of the envelope ends when the planet consumes all the gas available on its feeding zone,
\item or it opens up a gap on its orbit,
\item and finally the gas accretion ends when $\frac{dM_g}{dt} > \frac{1M_{\oplus}}{100 years}$. 
\end{itemize}
 
\subsection{Planetary Migration}\label{migracion}

The gravitational interaction between the embryos and the gaseous protoplanetary disc leads to an angular momentum exchange between them and as a consequence an orbital motion or migration of the embedded embryos occur \citep{b22}. For low mass planets the angular momentum flux injected into the disc is negligible when compared to the viscous transport of angular momentum, in this case we have the type I migration regime, following \citet{b23} the migration rate is,

\begin{equation}\label{migI}
\bigg(\frac{da}{dt}\bigg)_{migI}= c_{migI}[2.7+1.1\beta] \bigg(\frac{M_t}{M_{\star}}\bigg) \frac{\Sigma_g\, a^2}{M_{\star}}\bigg(\frac{a\Omega_K}{c_s}\bigg)^2 a\, \Omega_K 
\end{equation}  
as the time scale for type I migration is inversely proportional to the mass of the disc and the planet, it can be much shorter than the disc lifetime, so the factor $c_{migI}$ is introduced for considering effects that might slow down or even stop migration without introducing a mayor degree of complexity to the model and $\beta$ has the form,
\begin{equation}
\beta=-\frac{d\, log(\Sigma_g)}{d\, log(a)}=\gamma+(2-\gamma)\bigg(\frac{a}{a_c}\bigg)^{2-\gamma}
\end{equation}

When the planet reaches the mass necessary to open up a gap on its orbit, the angular momentum flux from the planet locally dominates the viscous flux and a  new regime of planetary migration begins. This is the type II regime, characterised by the next migration rate \citep{b24,b25},  

\begin{displaymath}
\bigg(\frac{da}{dt}\bigg)_{migII}\simeq 3sign(a-R_m)\alpha \frac{\Sigma_g(R_m)R_m^2}{M_t}\, \frac{\Omega_K(R_m)}{\Omega_K}
\end{displaymath}
\begin{equation}
\bigg(\frac{h(R_m)}{a}\bigg)^2\, a\Omega_K(R_m) 
\end{equation}  
with $\alpha=10^{-3}$ a dimensionless parameter which characterises the viscosity, $R_m=10 e^{\frac{2t}{\tau_{disc}}}~au$ and $\tau_{disc}$ the disc depletion time-scale.

We assume that both migration mechanisms stop when the core reaches the inner edge of the disc.

\section{Results}\label{resultados}

We perform a series of simulations varying the density profile and the type I migration rate and analyze which are the consequences on the planetary formation of considering a different prescription for the protoplanetary nebula profile and migration rates. In each simulation we generate 1000 discs. Each system evolves for 20 millions of years and the initial conditions for each one are chosen random taking into account the following conditions:

\begin{itemize}
\item{}The time-scale for the depletion of the gas, has a uniform log distribution between $10^6$ and $10^7$ years.
\item{}The stellar mass has a uniform distribution in log scale in the range of $0.7-1.4M_{\odot}$.
\item{}The distribution of metallicities of solar-like stars in the solar neighborhood follows a Gaussian distribution with $\mu=-0.02$ and dispersion 0.22 \citep{b21}.
\item{}The total mass of the disc is well approximated by a log-Gaussian distribution with mean $-2.05$ and dispersion $0.85$. We obtained this value by assuming a log-Gaussian distribution and performed a non-linear least square fit to the sample observed by \citet{b2} and \citet{b3}.
\item{}The characteristic radius, $a_c$ is also well approximated by a log-Gaussian distribution with $\mu=3.8$ and $\sigma=0.18$. This distribution was obtained with the same procedure described in the previous item.
\end{itemize}

\begin{figure}
  \begin{center}
    \subfigure[]{\label{nini-gama15}\includegraphics[angle=270,width=.5\textwidth]{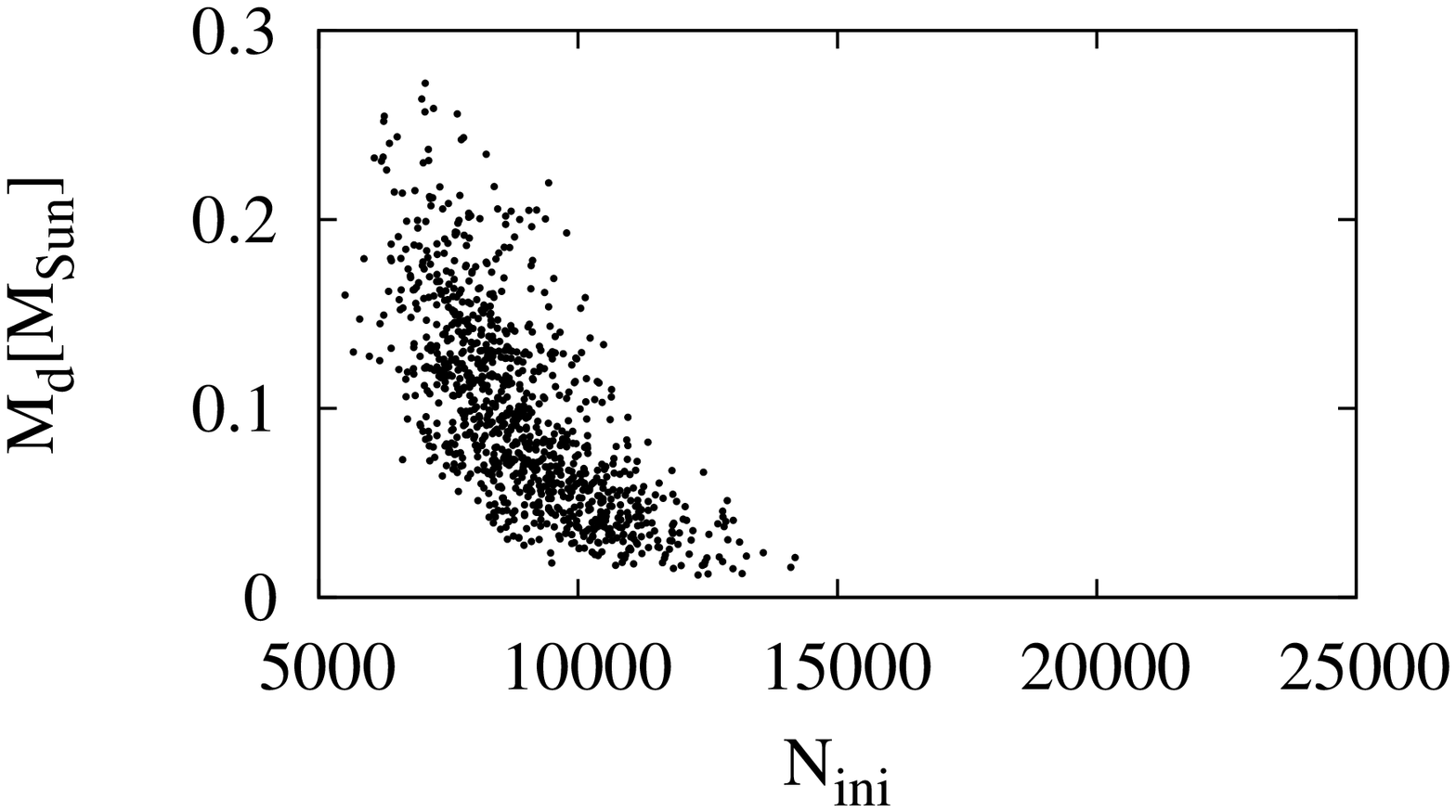}}
    \subfigure[]{\label{nini-gama1}\includegraphics[angle=270,width=.5\textwidth]{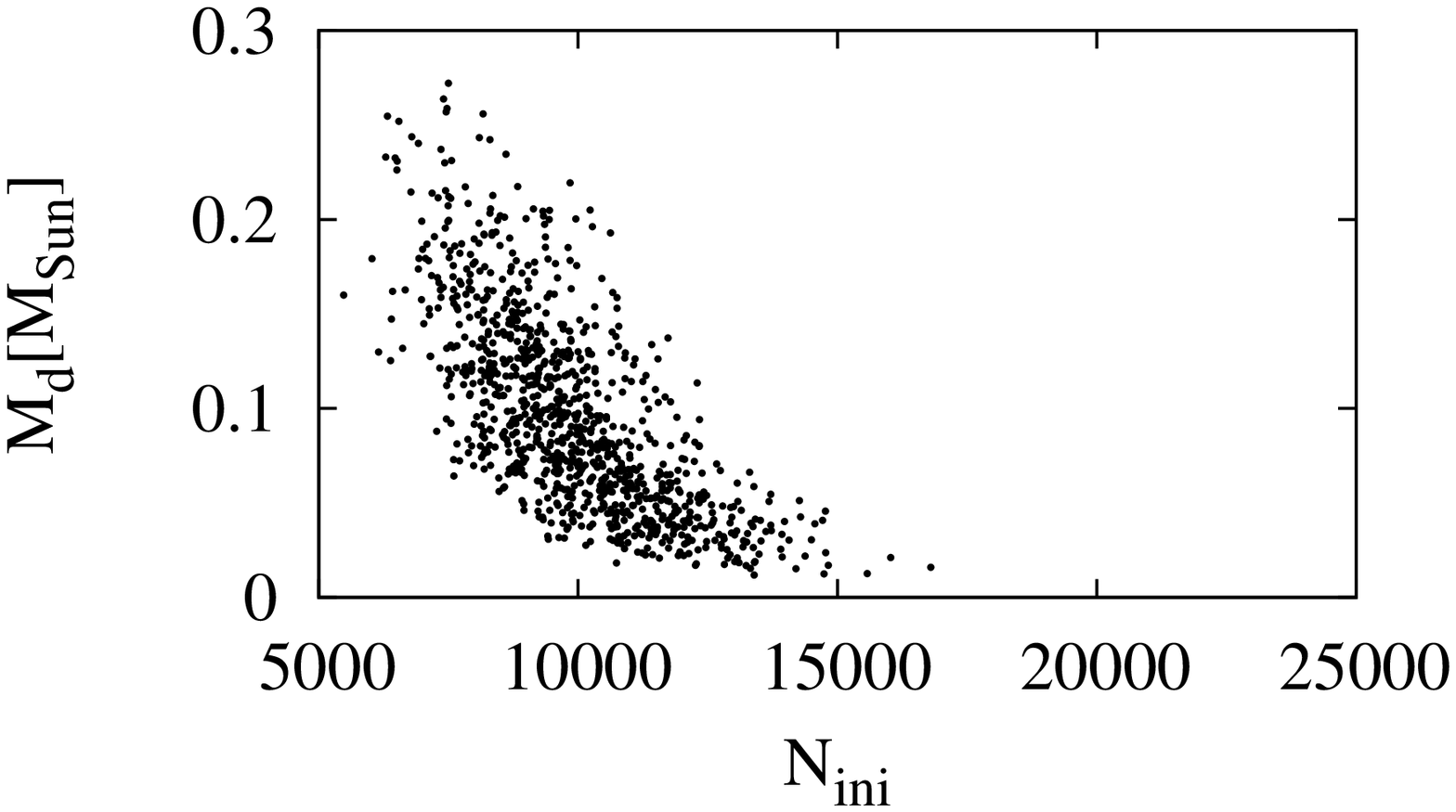}}
    \subfigure[]{\label{nini-gama05}\includegraphics[angle=270,width=.5\textwidth]{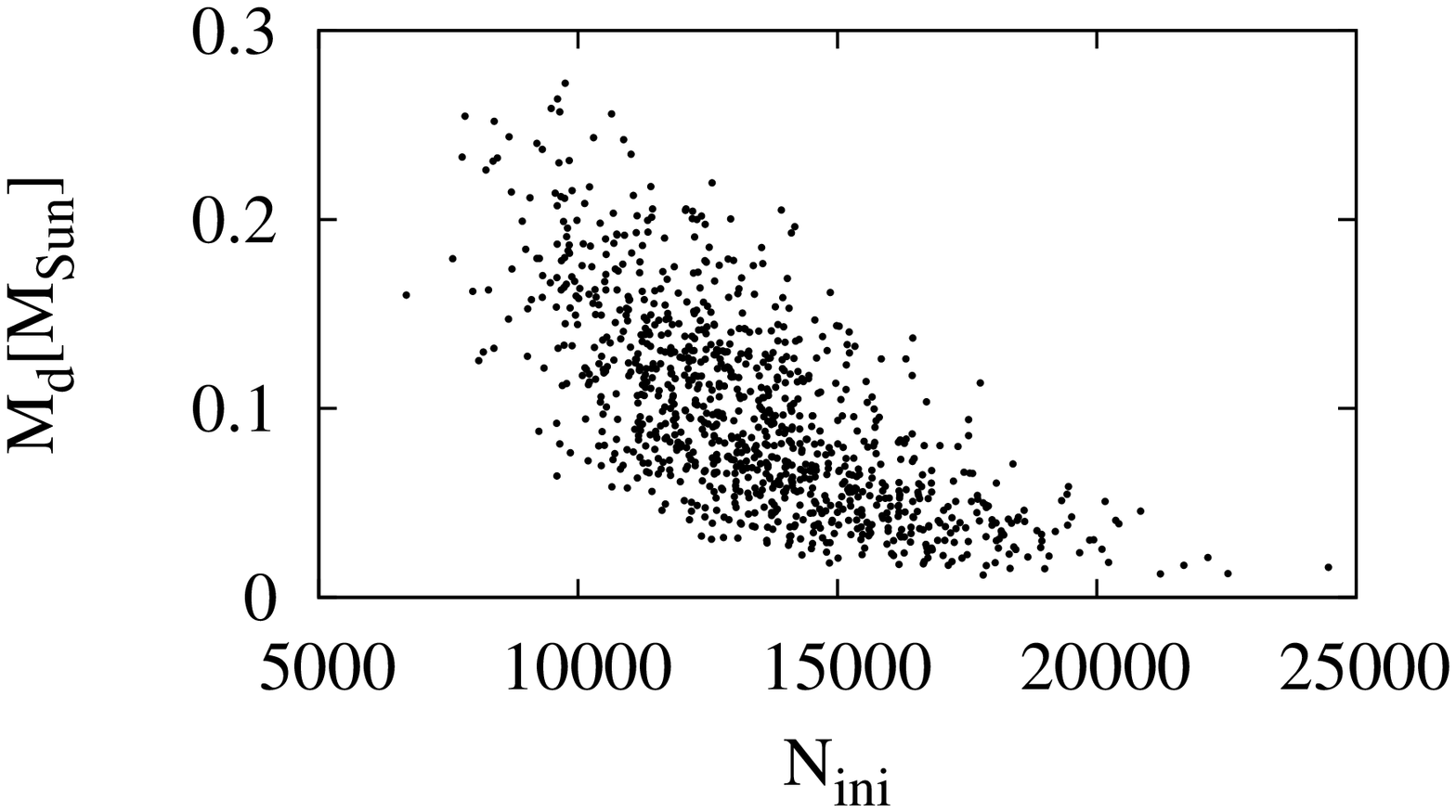}}
  \end{center}
  \caption{Initial number of embryos per disc as a function of the disc mass. The first figure (figure \ref{nini-gama05}) shows the initial number of embryos when the discs have a profile characterized by $\gamma=1.5$, when $\gamma=1$ the results are plotted in figure \ref{nini-gama1} and when $\gamma=0.5$ the initial number of embryos per disc are shown in figure \ref{nini-gama05}.}
  \label{Ninicial}
\end{figure}

We also note that the initial number of planets per disc depends on the disc size and mass as well as on the inner disc radius and the stellar mass, as is shown in Figure \ref{Ninicial}, where the initial number of embryos is plotted as a function of the disc mass for the three different values of $\gamma$ assumed in this work. Those discs characterized by a profile with $\gamma=1.5$ are shown in figure \ref{nini-gama15}, when $\gamma=1$ the results are shown in figure \ref{nini-gama1} and figure \ref{nini-gama05} represent the results when $\gamma=0.5$.  

We note in the figures, that the greater the mass of the disc, the lower the initial number of embryos. This is because the separation between the embryos is greater as the larger the initial mass of the embryo. On the other hand the initial mass of the embryo is larger when the density of solids is higher and this in turn, is higher as the greater the mass of the disc is. In summary, the greater the mass of the disc, the greater the separation between the embryos and consequently there will be a lower initial number of planetary cores. 

\begin{figure}
  \begin{center}
    \subfigure[]{\label{1ma-sinmig}\includegraphics[angle=270,width=.48\textwidth]{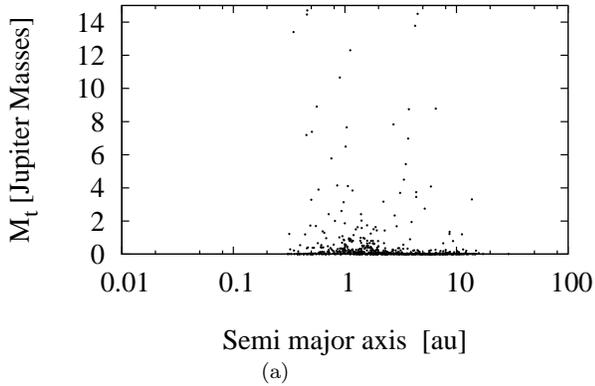}}
    \subfigure[]{\label{2ma-sinmig}\includegraphics[angle=270,width=.48\textwidth]{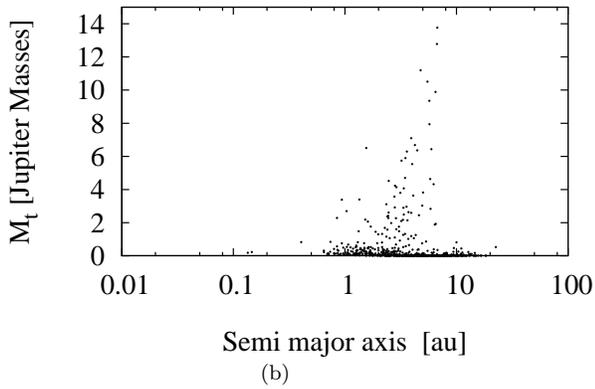}}
    \subfigure[]{\label{3ma-sinmig}\includegraphics[angle=270,width=.48\textwidth]{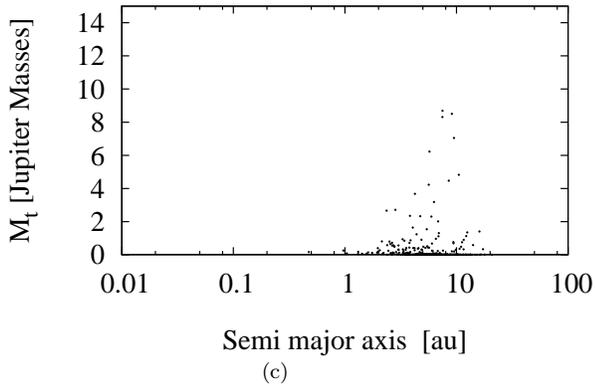}}
  \end{center}
  \caption{Mass and semi-major axis of all the planets formed in 1000 planetary systems, where the planetary migration was not considered. Figure \ref{1ma-sinmig} shows the results when we consider $\gamma=1.5$, figure \ref{2ma-sinmig} shows the mass and location of the planets when $\gamma= 1$ and finally the results obtained with $\gamma=0.5$ are shown in figure \ref{3ma-sinmig}.}
  \label{ma-sinmig}
\end{figure}

Figures \ref{ma-sinmig} show the mass and semi major axis distribution of all the planets that were formed in 1000 planetary systems, where the effect of planetary migration is not considered and the initial density disc profile is $\gamma=1.5$ in \ref{1ma-sinmig}, $\gamma=1$ in \ref{2ma-sinmig} and $\gamma=0.5$ in the last figure.  The first figure shows the results obtained with the profile more similar {\bf to} the MMSN model, where the solids are available just beyond the snow line, fact that favours the formation of giant planets at this location as we can see in the figure. This profile is also the steepest, making more feasible the formation of a single giant planet close to the snow line. In the second case, when $\gamma=1$ the profile is a bit soften, and as a consequence there are more solids available in the outer parts of the disc, fact that promotes the formation of giant planets further  from the central star. Finally in the last case there are giant planets at around $15$~au and even further. 

\textit{As a result it is noticed that in order to form giant planets as the ones observed in our own Solar System in their current locations, it is more convenient to consider a protoplanetary nebula model with a softer profile.}

\begin{figure}
  \begin{center}
    \subfigure[]{\label{1ma-c001}\includegraphics[angle=270,width=.48\textwidth]{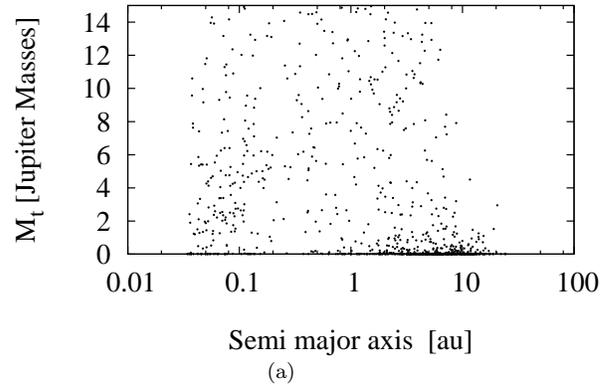}}
    \subfigure[]{\label{2ma-c001}\includegraphics[angle=270,width=.48\textwidth]{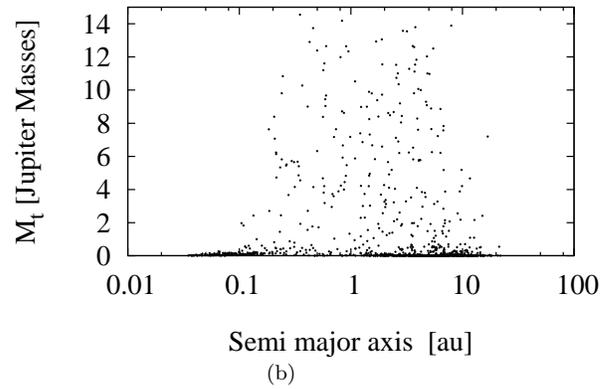}}
    \subfigure[]{\label{3ma-c001}\includegraphics[angle=270,width=.48\textwidth]{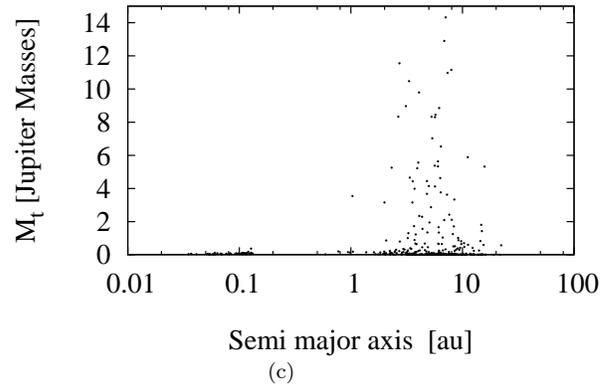}}
  \end{center}
  \caption{The figures show the distribution of mass and semi-major axis of all the planets formed when the planetary migration is considered but the type I is delayed 100 times. The first figure shows the planets formed when we consider $\gamma=1.5$, figure \ref{2ma-c001} shows the results obtained with $gamma= 1$ and figure  \ref{3ma-c001} shows the distribution obtained with $\gamma=0.5$.}
  \label{ma-c001}
\end{figure}

The next set of plots (figures \ref{ma-c001}), are the results when we do consider planetary migration, but the type I planetary migration is delayed 100 times ($c_{migI}=0.01$). Figure \ref{1ma-c001} shows the distribution when the protoplanetary nebula was modeled with a density profile with $\gamma=1.5$, figure \ref{2ma-c001} represents all the planets formed when $\gamma=1$ and when $\gamma=0.5$ we found the distribution observed in \ref{3ma-c001}.

Here we still note that the giant planet formation is favoured when a softer profile is considered. One consequence of the planetary migration is that a larger population of giant planets is observed. This is due to the fact that when a planet is migrating \textit{slowly} it finds more solids available to accrete in the new feeding zone, and it grows faster. We also note that the more massive planets are found when $\gamma=1.5$, this is because this profile overestimates the mass of solids just beyond the snow line and the giant planets formed there have more solids available than in the other cases and as a result became bigger. Another point is that a population of giant planets closer to the star begins to be observed, fact that was not seen when the migration was not considered. This population of Hot-Jupiters will be called hereafter population I. 

\begin{figure}
  \begin{center}
    \subfigure[]{\label{1ma-c01}\includegraphics[angle=270,width=.48\textwidth]{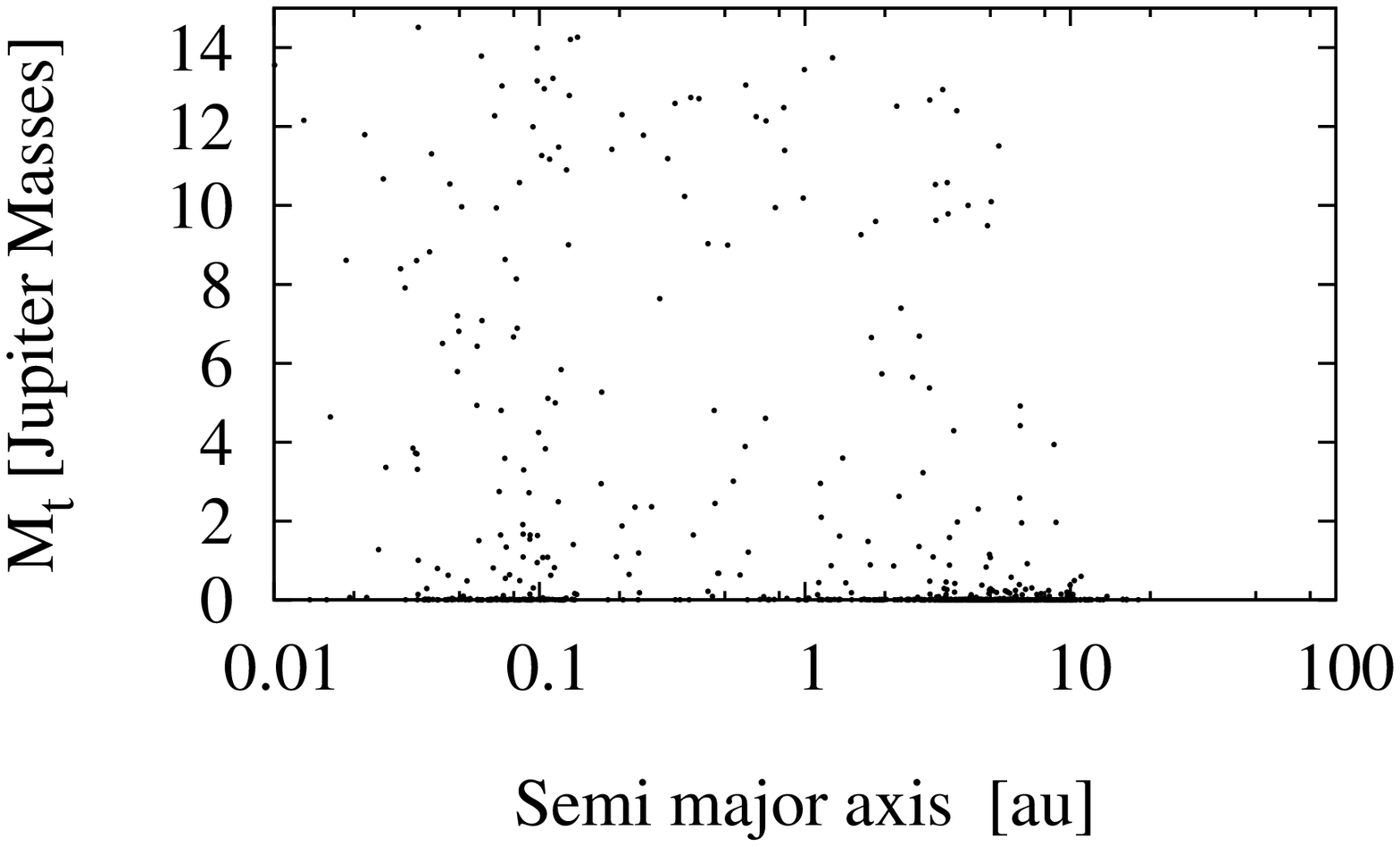}}
    \subfigure[]{\label{2ma-c01}\includegraphics[angle=270,width=.48\textwidth]{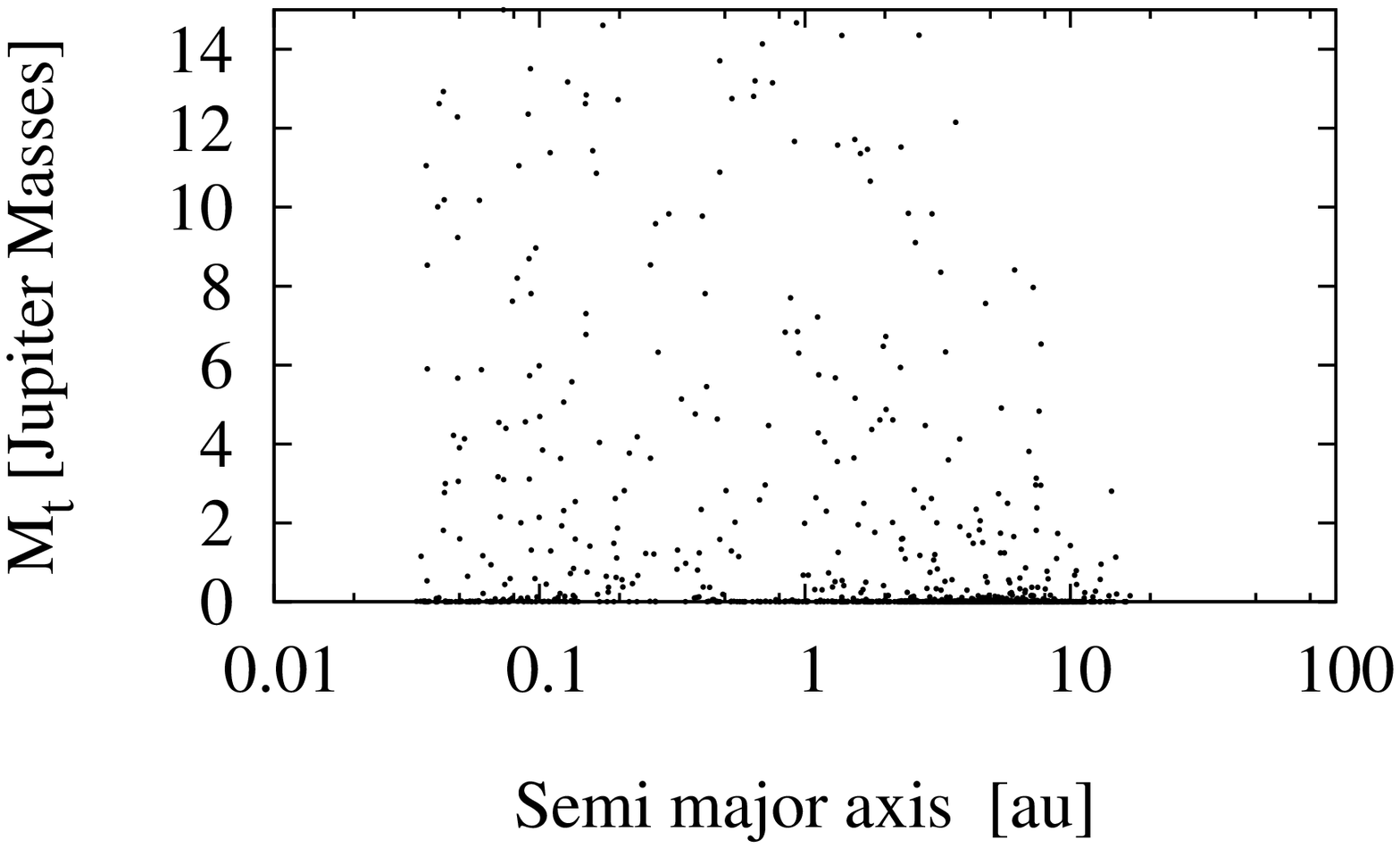}}
    \subfigure[]{\label{3ma-c01}\includegraphics[angle=270,width=.48\textwidth]{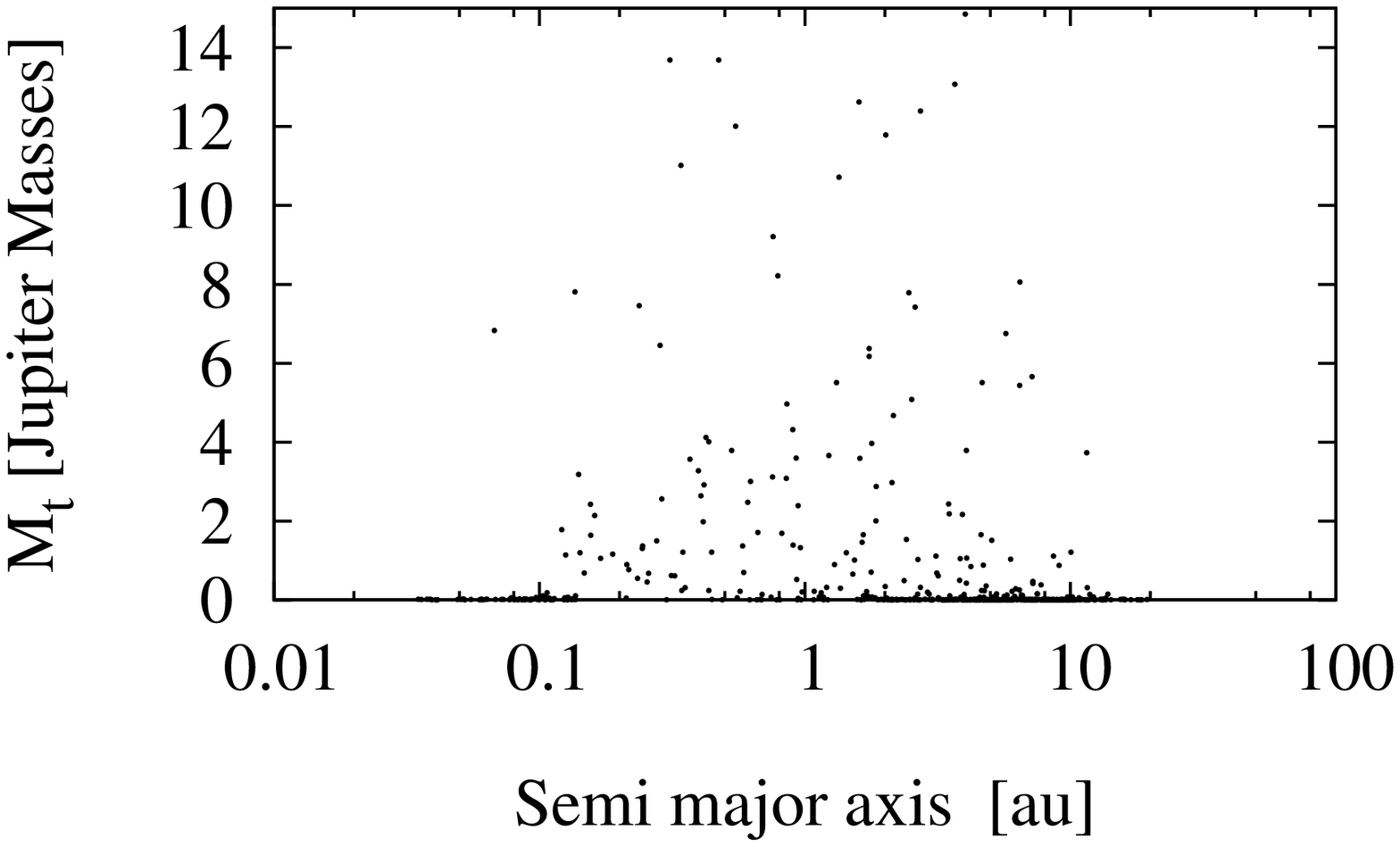}}
  \end{center}
  \caption{Mass and semi-major axis of all the planets formed in 1000 planetary systems when the planetary migration was delayed 10 times. Figure \ref{1ma-c01} shows the results when we consider $\gamma=1.5$, figure \ref{2ma-c01} shows the mass and location of the planets when $\gamma= 1$ and finally the results obtained with $\gamma=0.5$ are shown in figure \ref{3ma-c01}.}
  \label{ma-c01}
\end{figure}

If the migration is delayed only 10 times ($c_{mig}=0.1$) we found the distribution shown in figure \ref{ma-c01}. Figure \ref{1ma-c01} shows the mass and semi major axis when $\gamma=1.5$, in figure \ref{2ma-c01} the profile is bit softer ($\gamma=1$), and the results with $\gamma=0.5$ are shown in figure \ref{3ma-c01}.      

Although migration is very fast, still can be seen that the smaller $\gamma$ is, the greater the chance of forming giant planets farther away from the central star. In these distributions there are a more marked differentiation between two populations of giant planets that had begun to appear in figures \ref{ma-c001}: the population I of giant planets close to the star, and another one of giant planets located beyond the ice line, farther of the host star. Since this differentiation is not observed in the distribution found without migration (figures \ref{ma-sinmig}), this must be an effect caused by the orbital motion of the embryos. Those planets that were initially located in a region reached in solids grow faster and become giants in a time scale shorter than the type I migration time scale and starts to migrate with the slower type II migration regime, as a result, they remain close to the zone where they were originally formed. On the other hand we have the population of planets which where initially located inside the ice line, closer to the star, in this region the solids available are not enough to form giant planets faster than the type I time-scale, so they were moved rapidly inwards. On their path towards the star they found other embryos to collide with and new material to grow, so they became Hot-Jupiters. We note that when $\gamma=0.5$ the population is not well marked as in the other cases, this is because with this density profile, the giant planets are formed in the outer parts of the disc and there are no material available in the inner parts, so these discs do not favored the formation of hot Jupiters.

\begin{figure}
  \begin{center}
    \subfigure[]{\label{1ma-c1}\includegraphics[angle=270,width=.48\textwidth]{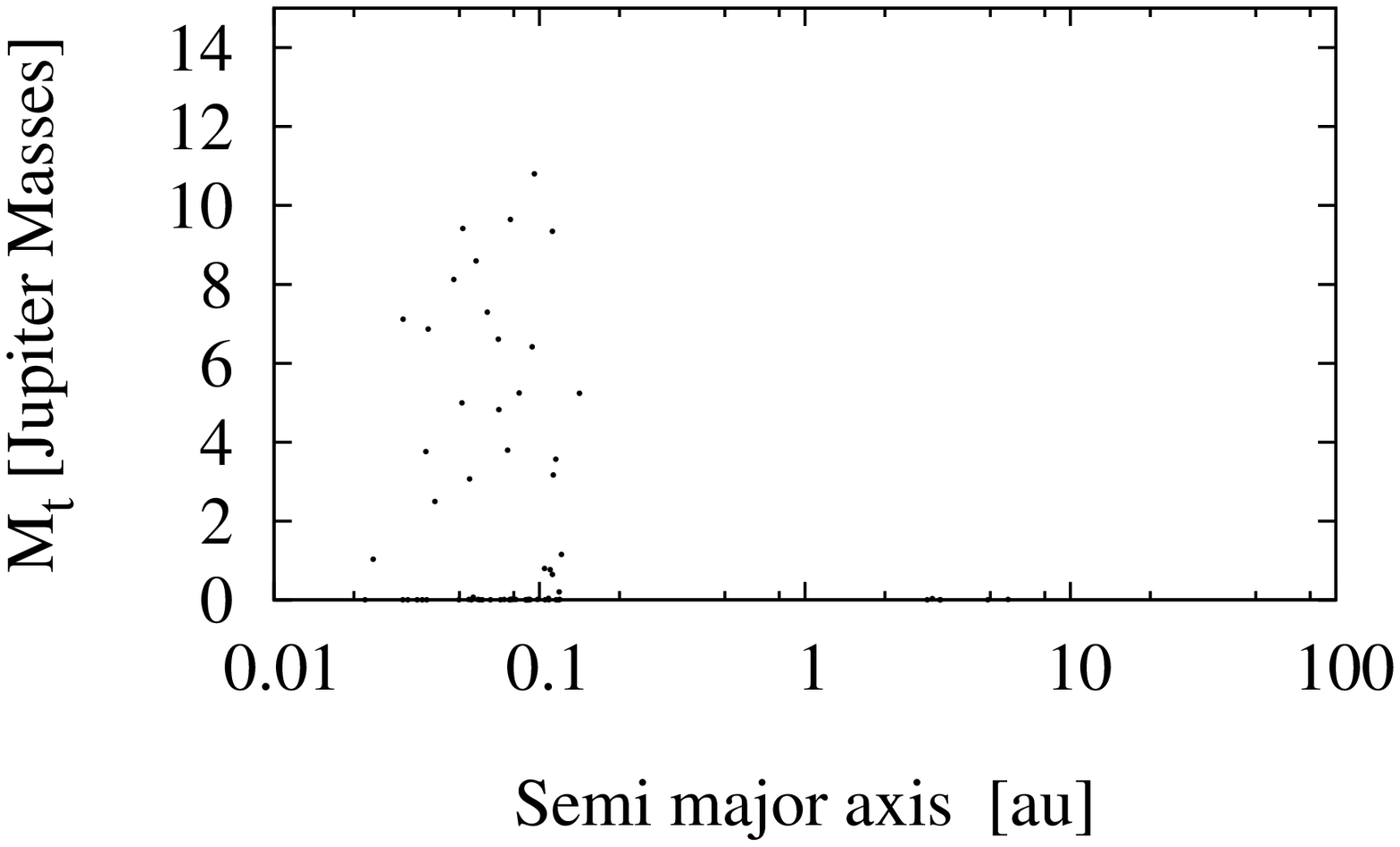}}
    \subfigure[]{\label{2ma-c1}\includegraphics[angle=270,width=.48\textwidth]{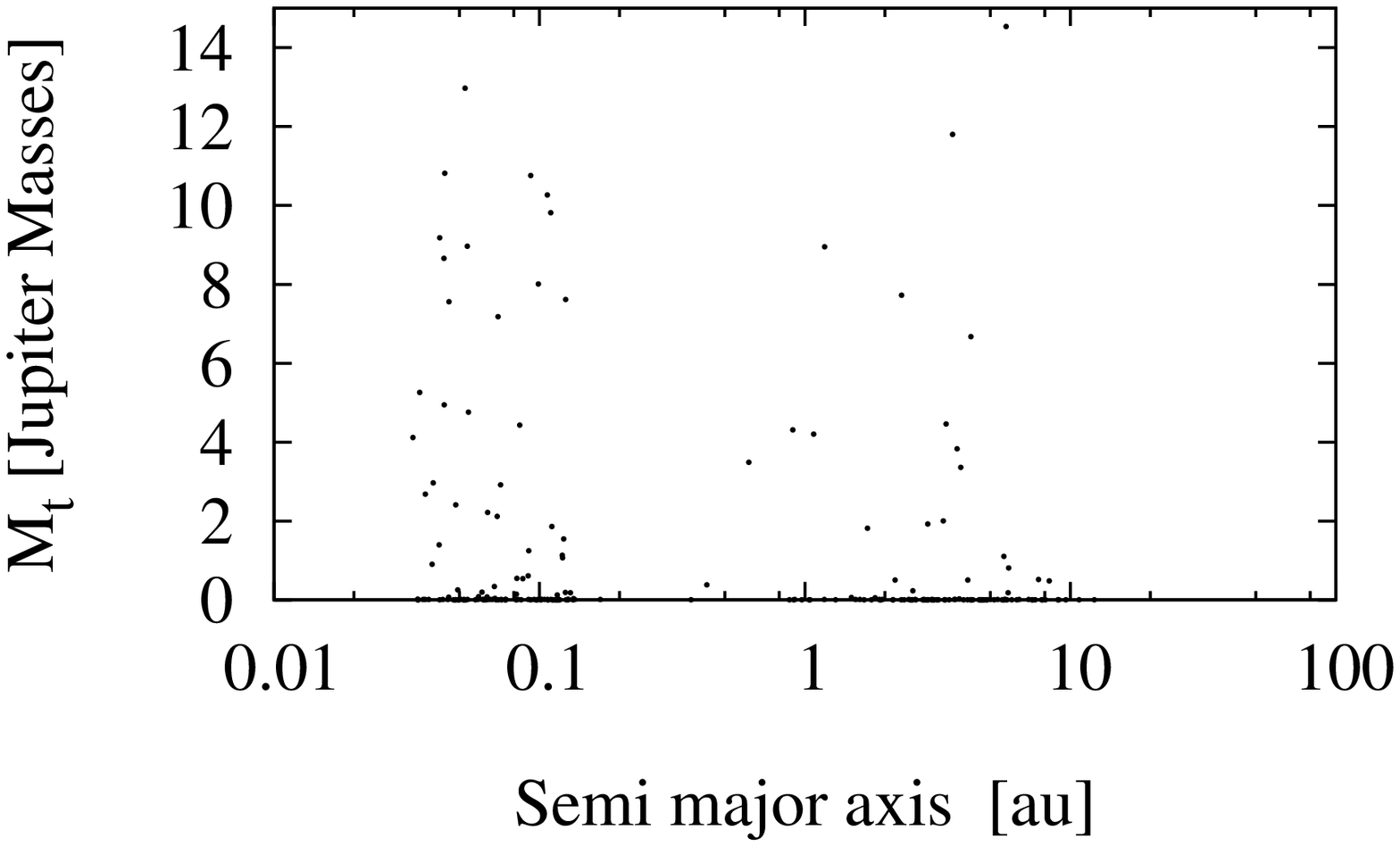}}
    \subfigure[]{\label{3ma-c1}\includegraphics[angle=270,width=.48\textwidth]{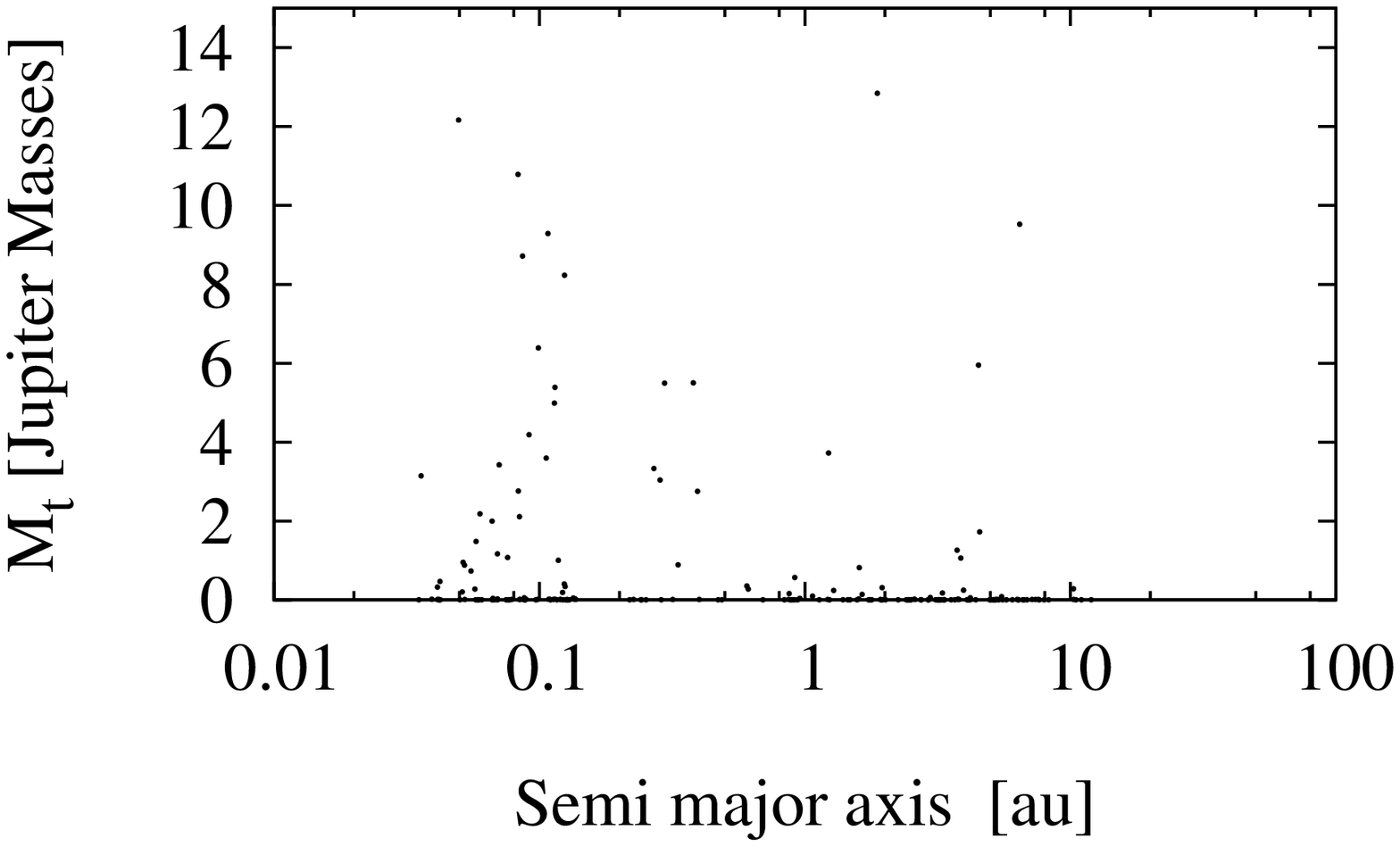}}
  \end{center}
  \caption{The figure shows the distribution of mass and semi-major axis found when we do not delayed planetary migration. When the density profiles corresponds to a value of $\gamma$ equal to 1.5 we observe figure \ref{1ma-c1}, figure \ref{2ma-c1} corresponds to $\gamma= 1$ and if we model the protoplanetary nebula with $\gamma=0.5$ we obtained the distribution observed in the last figure (Fig. \ref{3ma-c1}).}
  \label{ma-c1}
\end{figure}

Finally if we assume that there are no effects which act stopping or slowing down the migration rate ($c_{migI}=1$), we observe the distribution shown in figures \ref{1ma-c1} ($\gamma=1.5$), \ref{2ma-c1} ($\gamma=1$) and \ref{3ma-c1} ($\gamma=0.5$). In this cases the migration rate is faster than in the previous cases as a consequence some planets migrate really fast, faster than the depletion time scale, so the cores did not have enough time to grow and remain as small embryos. For this reason we note that the population of giant planets decreased in comparison with the cases of slower migration rate, and most of the planets are really close to the central star, located in the inner edge of the disc, specially when $\gamma=1.5$, because in this case the planets are formed closer to the star.

\subsection{Comparison with Exoplanets Observations}\label{comparo-obser}

In order to know which parameters must be considered to get a better representation of reality, in this section we compare the results obtained in our simulations with the mass and semi major axis distribution of observed exoplanets. 

\begin{figure}
  \begin{center}
    \includegraphics[angle=270,width=.5\textwidth]{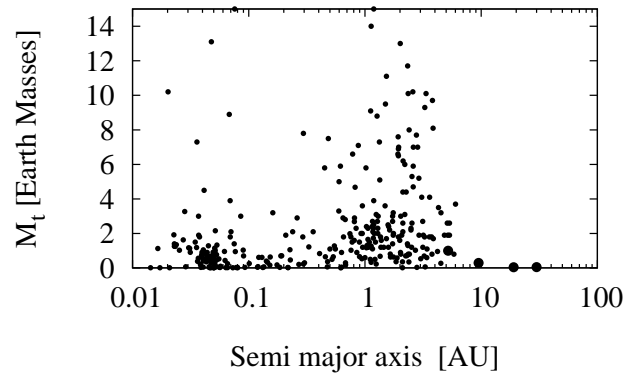}
  \end{center}
  \caption{In the Figure the current mass and semi major axis distribution of observed exoplanets is shown, where the giant planets in the Solar System are included(4 big dots) and the exoplanets found orbiting double or multiple stars were excluded of the sample.}
  \label{observaciones}
\end{figure}

Figure \ref{observaciones} shows the mass and semi major axis distribution of observed exoplanets \footnote{http://exoplanets.org} and the four giant planets in the Solar System, which are characterised for the biggest dots. We excluded of the sample those discovered planets that orbit binary and multiple stars, since the mechanisms of formation of these planets may not be explained with our model.

We clearly notice in the Figure two populations: one is the population I and the another one are those exoplanets centered about $\sim 3$~au (population II, hereafter).

Although in this work we are focused on the planetary migration as a possible explanation of these populations observed, it is not the only mechanism proposed when  trying to explain how hot Jupiters came to be so close to their parent star: Kozai cycles and planet scattering should provided us with an alternative explanation. These mechanisms excite the sky-projected obliquity of the planet and should provide us with a planet population on misaligned orbits with respect to their stars rotation, that it is now beginning to be detected. In the work of \citet{b49}, they present observations of some exoplanets who present this misalignments. So this is saying that the population I could be explained by a combination of mechanisms.

\begin{figure}
  \begin{center}
    \subfigure[]{\label{1comp-sinmig}\includegraphics[angle=270,width=.48\textwidth]{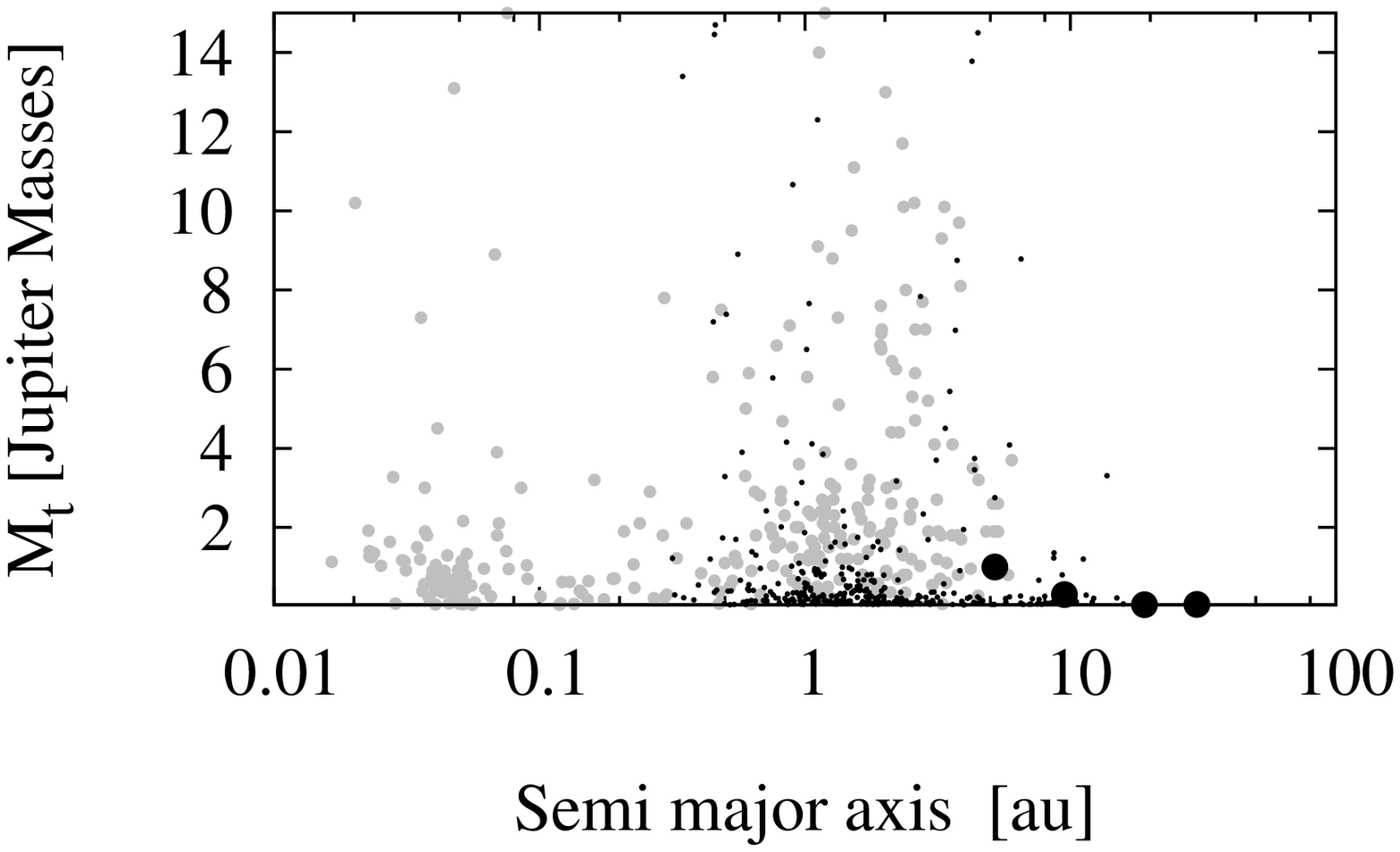}}
    \subfigure[]{\label{2comp-sinmig}\includegraphics[angle=270,width=.48\textwidth]{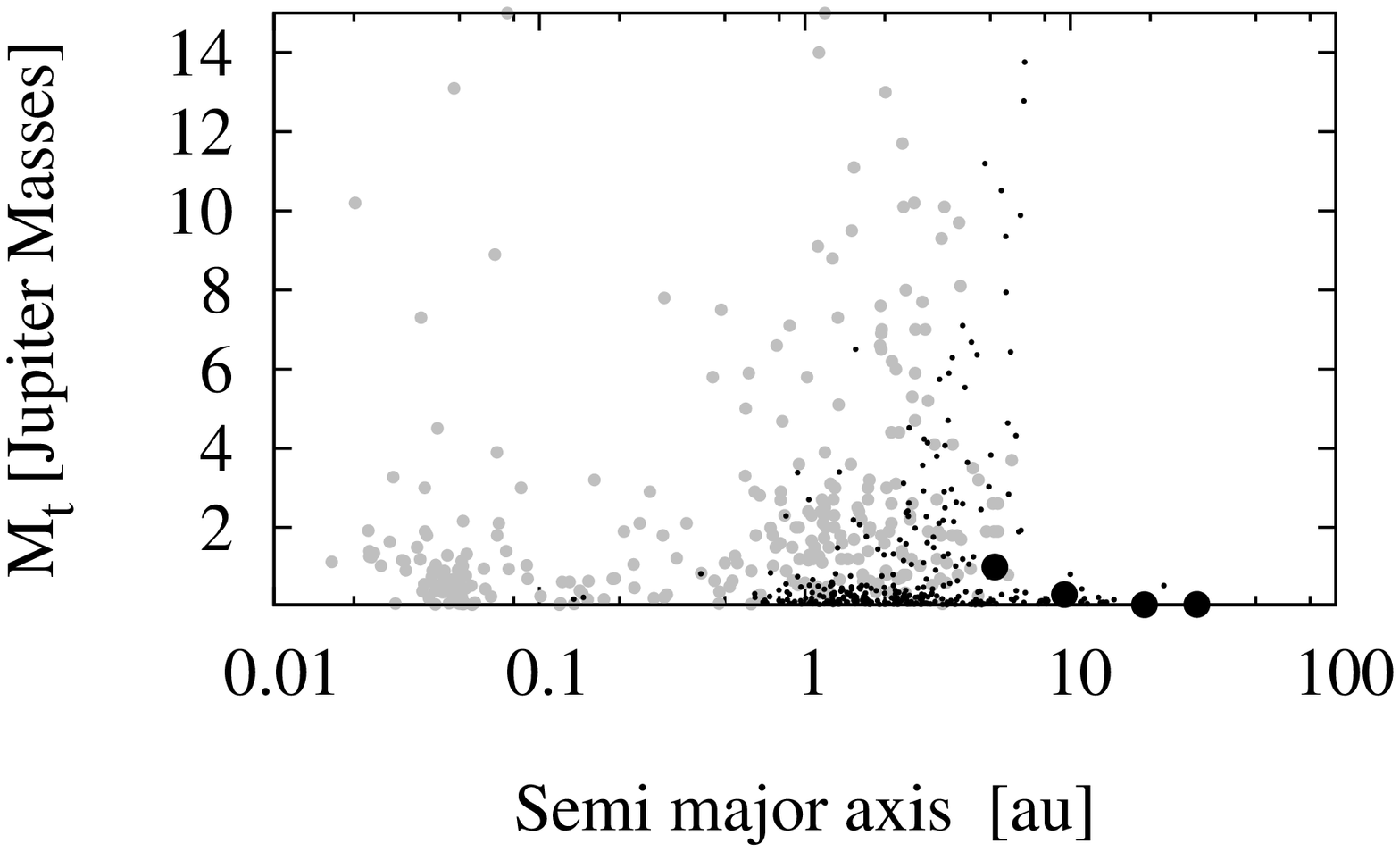}}
    \subfigure[]{\label{3comp-sinmig}\includegraphics[angle=270,width=.48\textwidth]{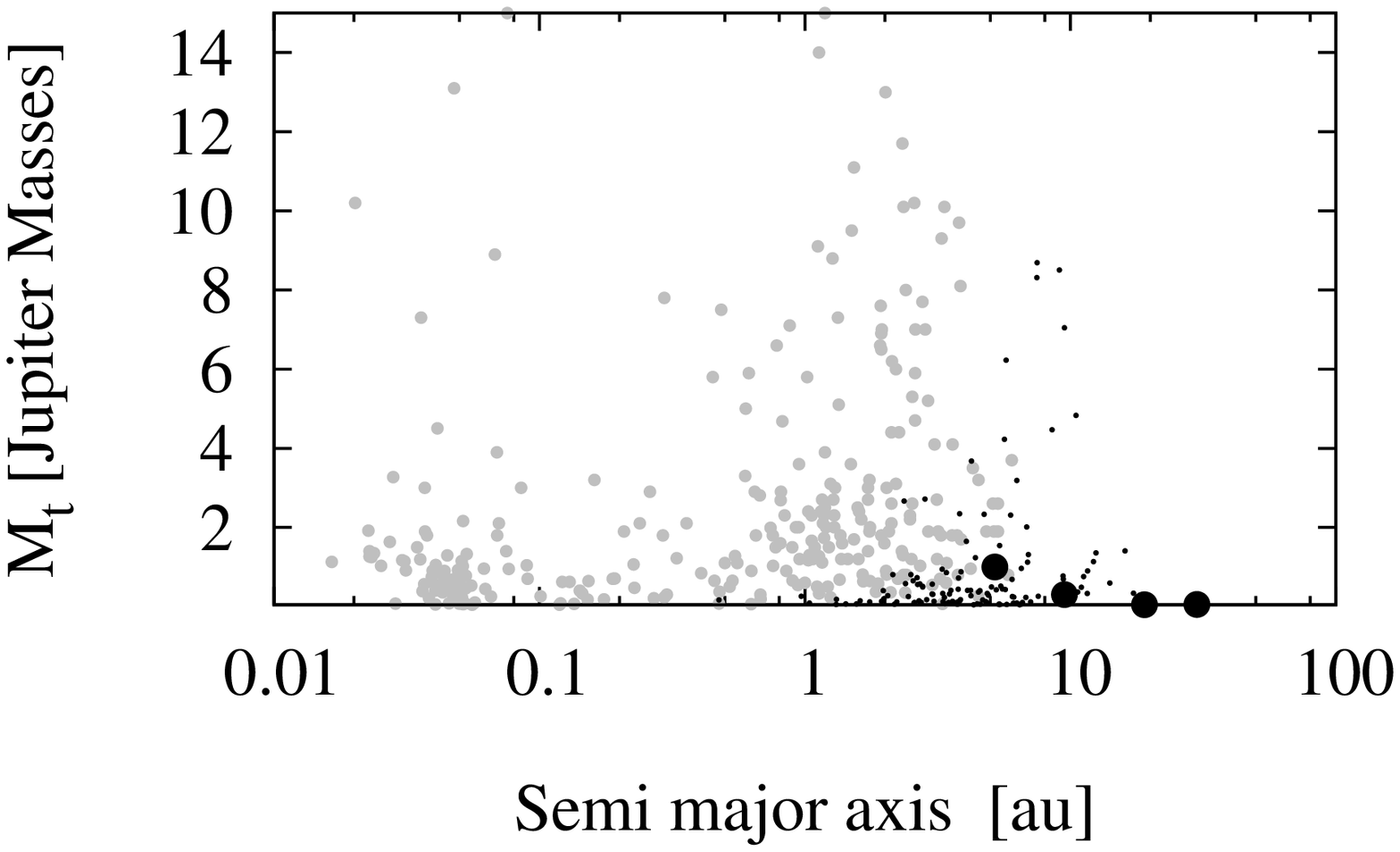}}
  \end{center}
  \caption{In the figure we show the mass and semi major axis distribution obtained when planetary migration is not considered which are the small black dots, the distribution of observed exoplanets are the grey dots and the giant planets in the Solar System (the big black dots). Figure \ref{1comp-sinmig} shows the results of the simulations when $\gamma=1.5$, \ref{2comp-sinmig} when $\gamma=1$ and the results with $\gamma)0.5$ are shown in \ref{3comp-sinmig}}
  \label{comp-sinmig}
\end{figure}

\begin{figure}
  \begin{center}
    \subfigure[]{\label{1comp-c001}\includegraphics[angle=270,width=.48\textwidth]{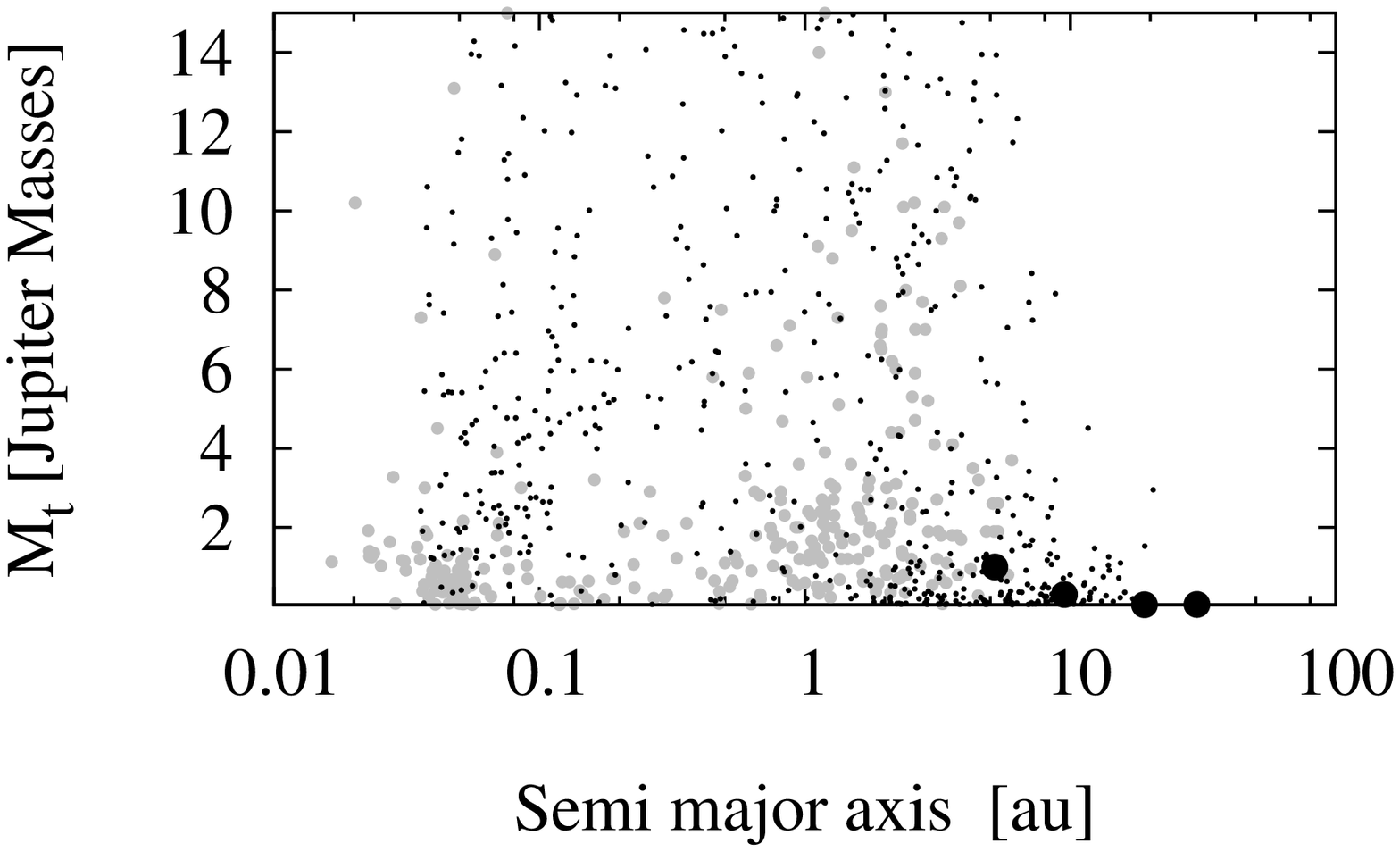}}
    \subfigure[]{\label{2comp-c001}\includegraphics[angle=270,width=.48\textwidth]{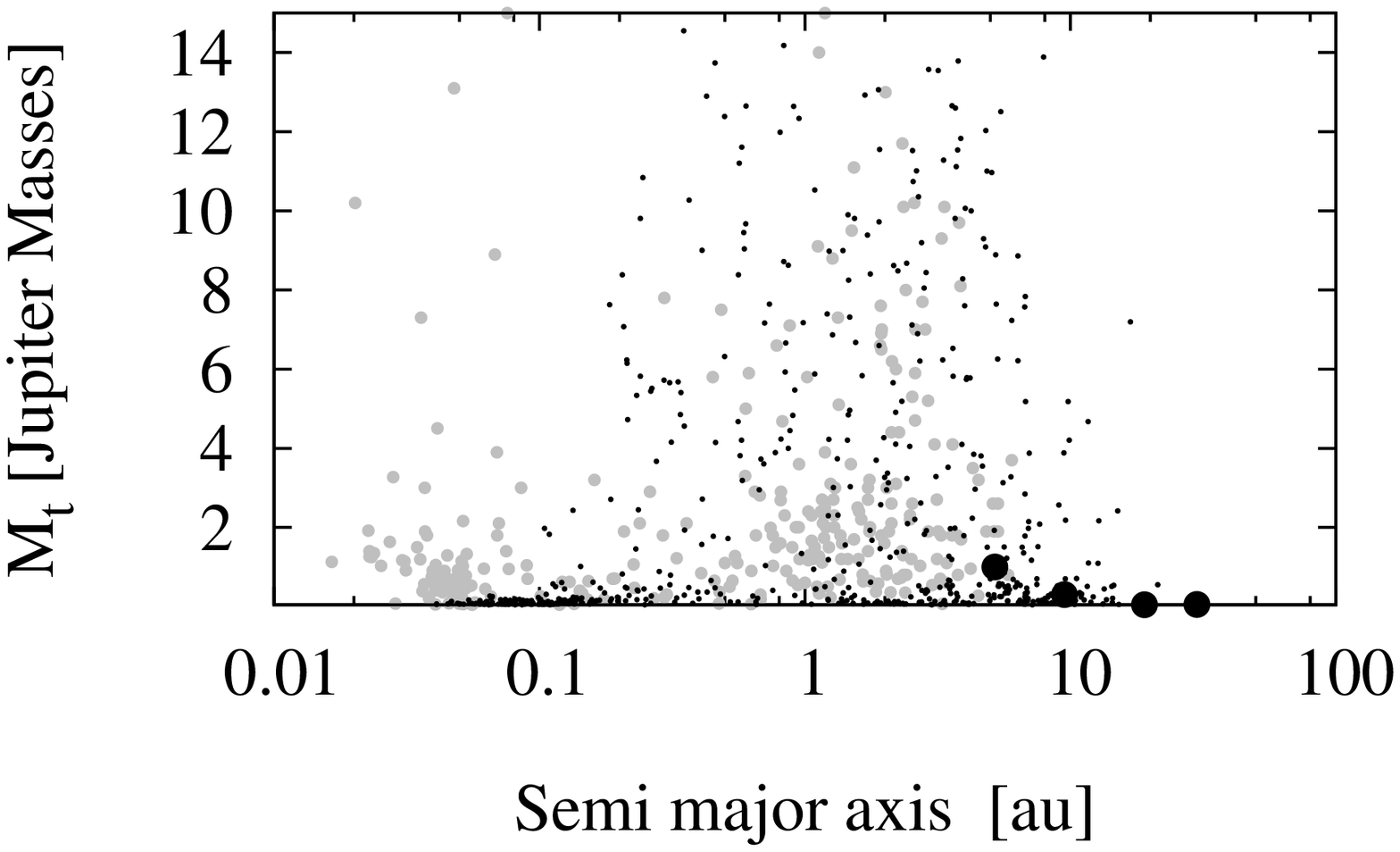}}
    \subfigure[]{\label{3comp-c001}\includegraphics[angle=270,width=.48\textwidth]{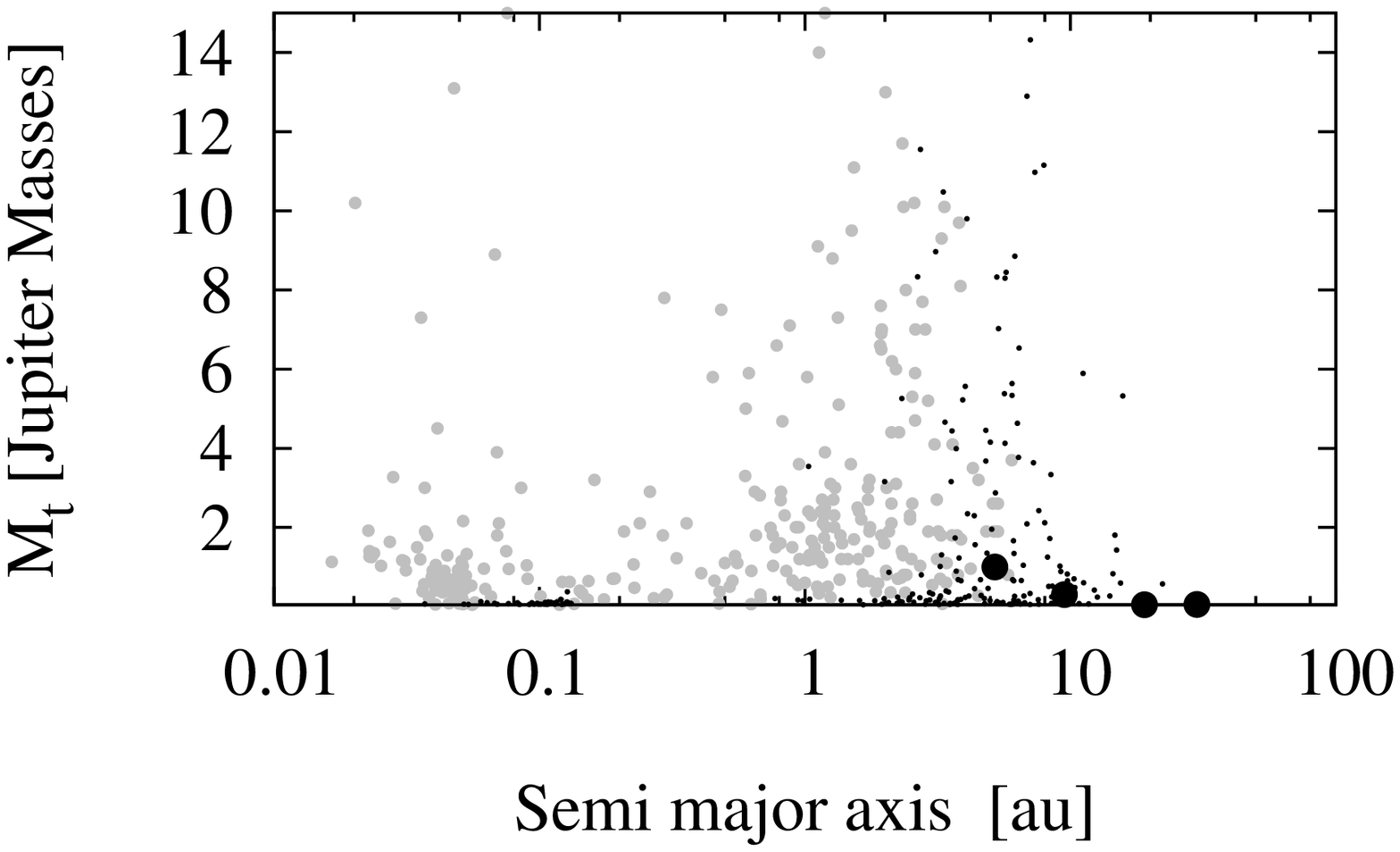}}
  \end{center}
  \caption{Mass and semi major axis distribution of the planets found in the simulation  when the migration is delayed 100 times are the small black dots, the current distribution of exoplants observed are the grey dots and the giant planets in the Solar System are the black big dots. Figure \ref{1comp-c001} shows the results of the simulations when $\gamma=1.5$, $\gamma=1$ are shown in \ref{2comp-c001} and the last figure shows the results with $\gamma=0.5$.}
  \label{comp-c001}
\end{figure}

\begin{figure}
  \begin{center}
    \subfigure[]{\label{1comp-c01}\includegraphics[angle=270,width=.48\textwidth]{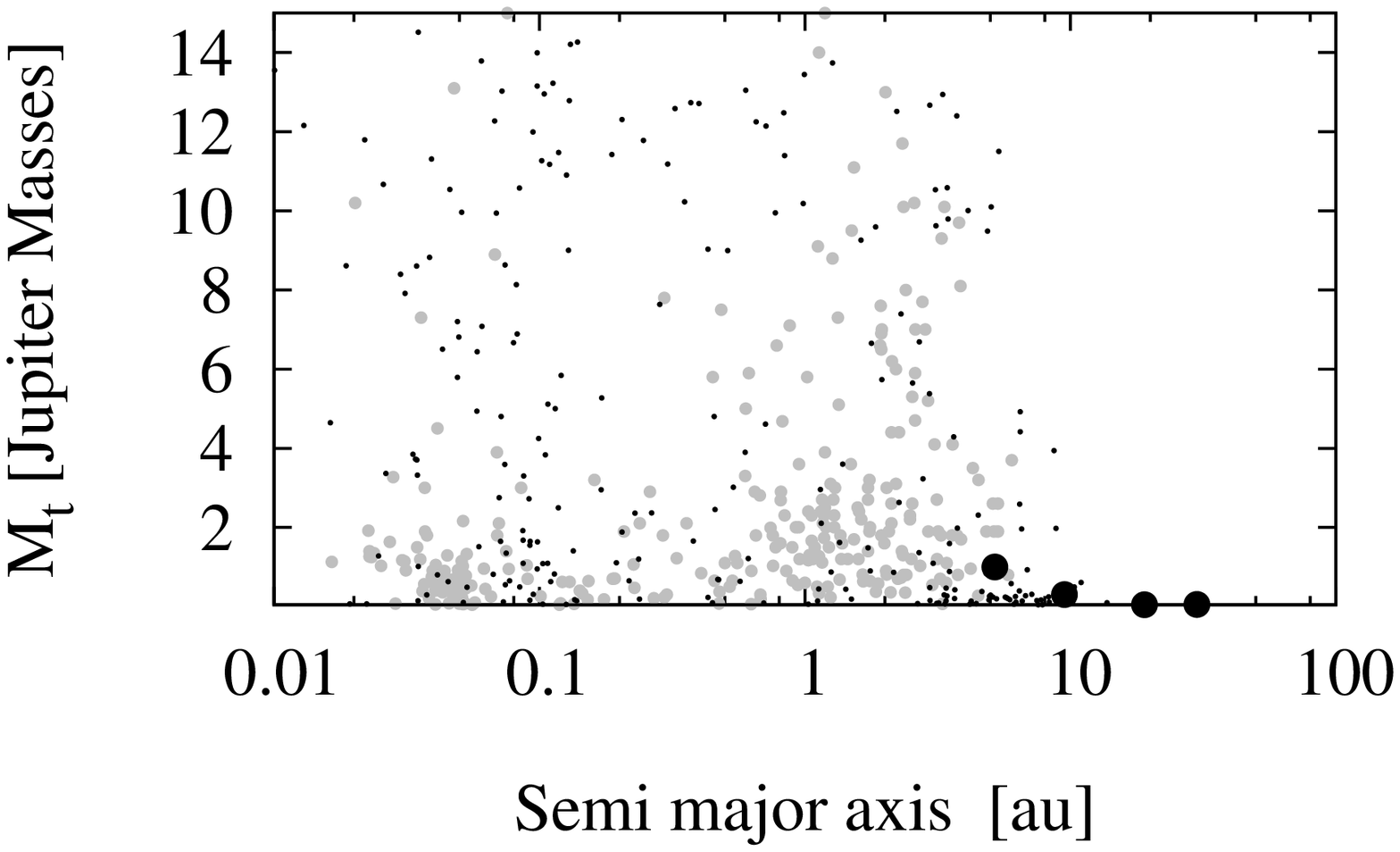}}
    \subfigure[]{\label{2comp-c01}\includegraphics[angle=270,width=.48\textwidth]{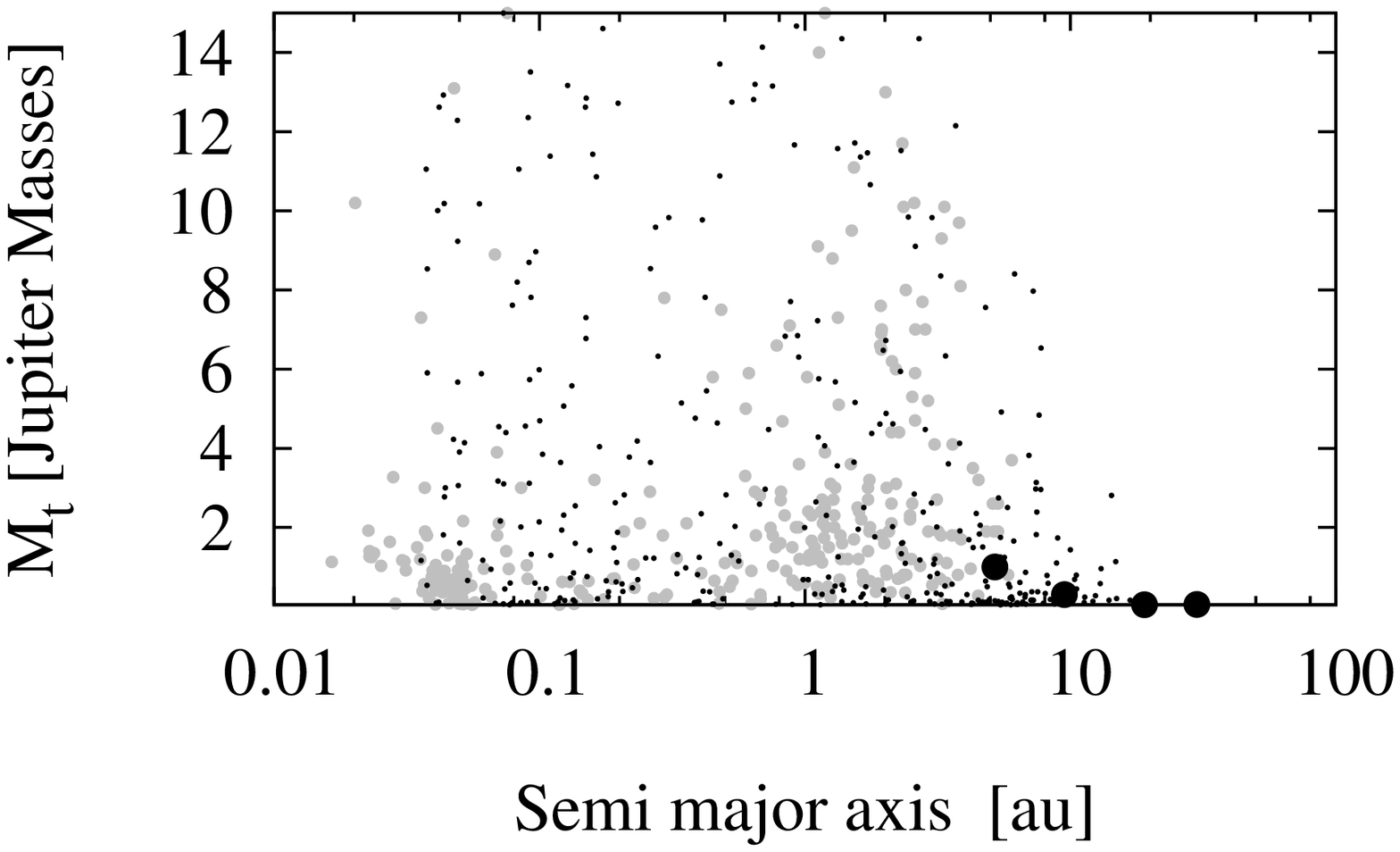}}
    \subfigure[]{\label{3comp-c01}\includegraphics[angle=270,width=.48\textwidth]{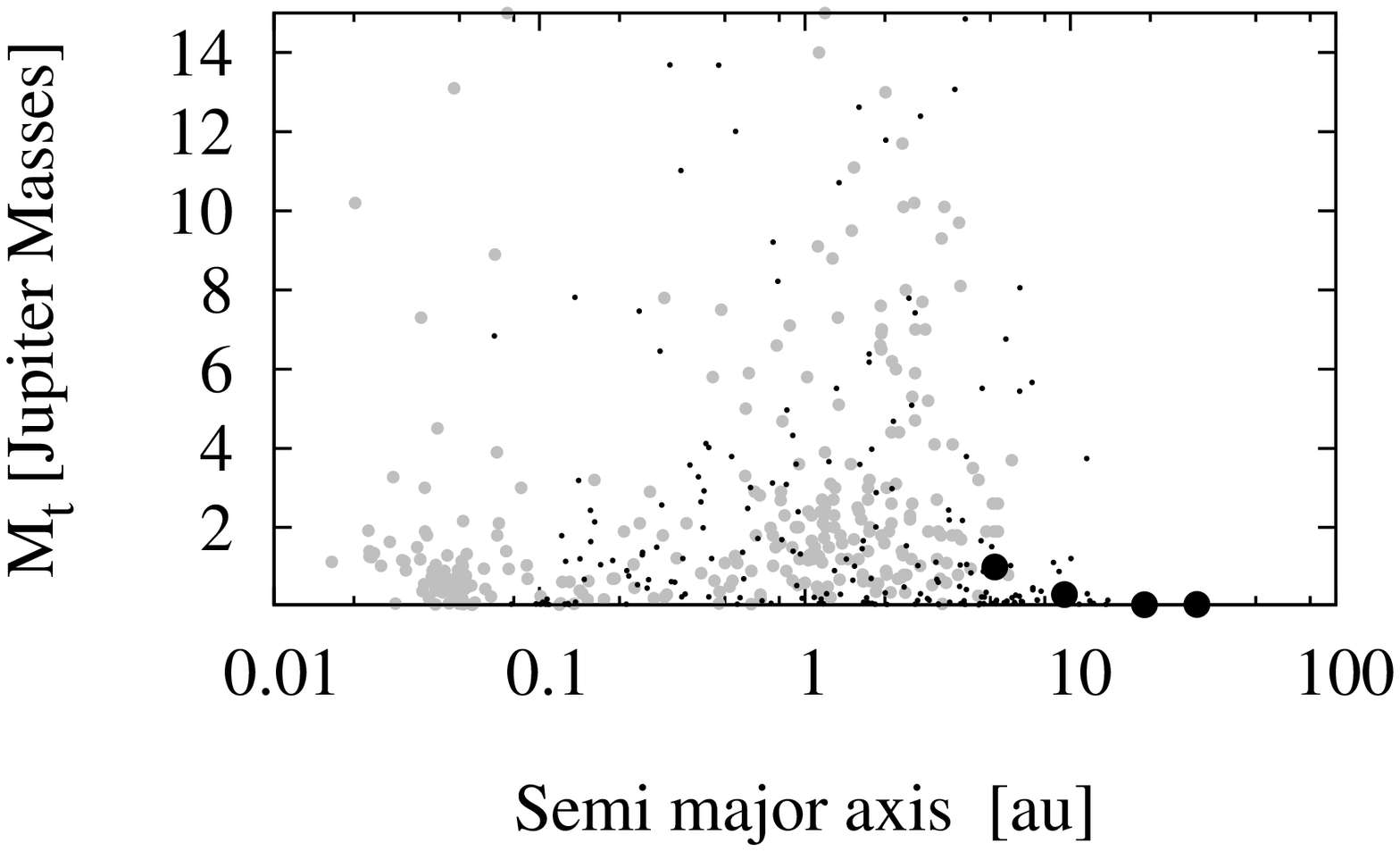}}
  \end{center}
  \caption{The figures show the mass and semi major axis of the giant planets in the Solar System, which are the biggest black dots, the distribution of observed extrasolar planets (grey dots), and the simulations results (the small black dots), when planetary migration is considered and delayed 10 times and $\gamma=1.5$ (\ref{1comp-c01}), $\gamma=1$ (\ref{2comp-c01}) and $\gamma=0.5$ (\ref{3comp-c01}).}
  \label{comp-c01}
\end{figure}

\begin{figure}
  \begin{center}
    \subfigure[]{\label{1comp-c1}\includegraphics[angle=270,width=.48\textwidth]{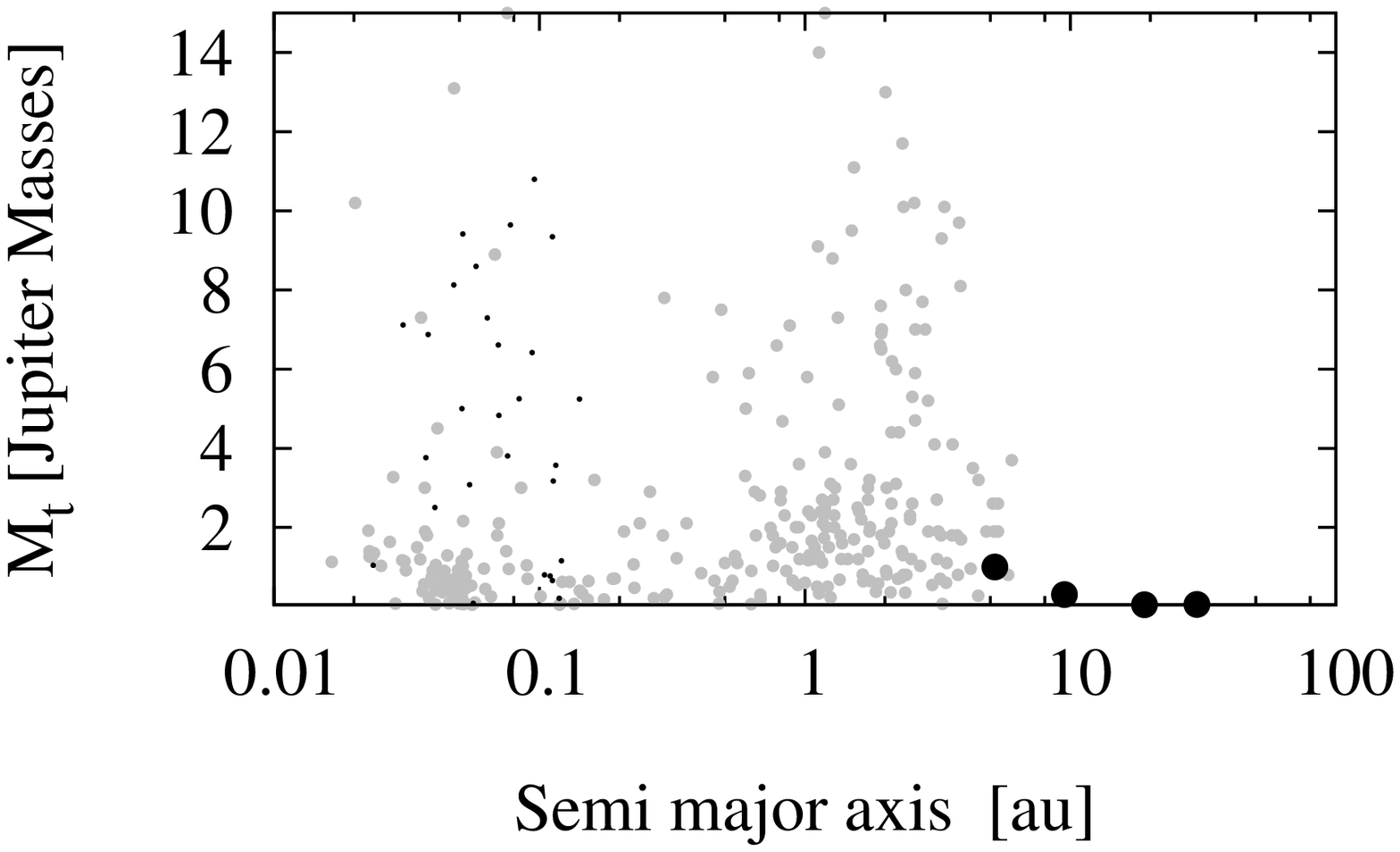}}
    \subfigure[]{\label{2comp-c1}\includegraphics[angle=270,width=.48\textwidth]{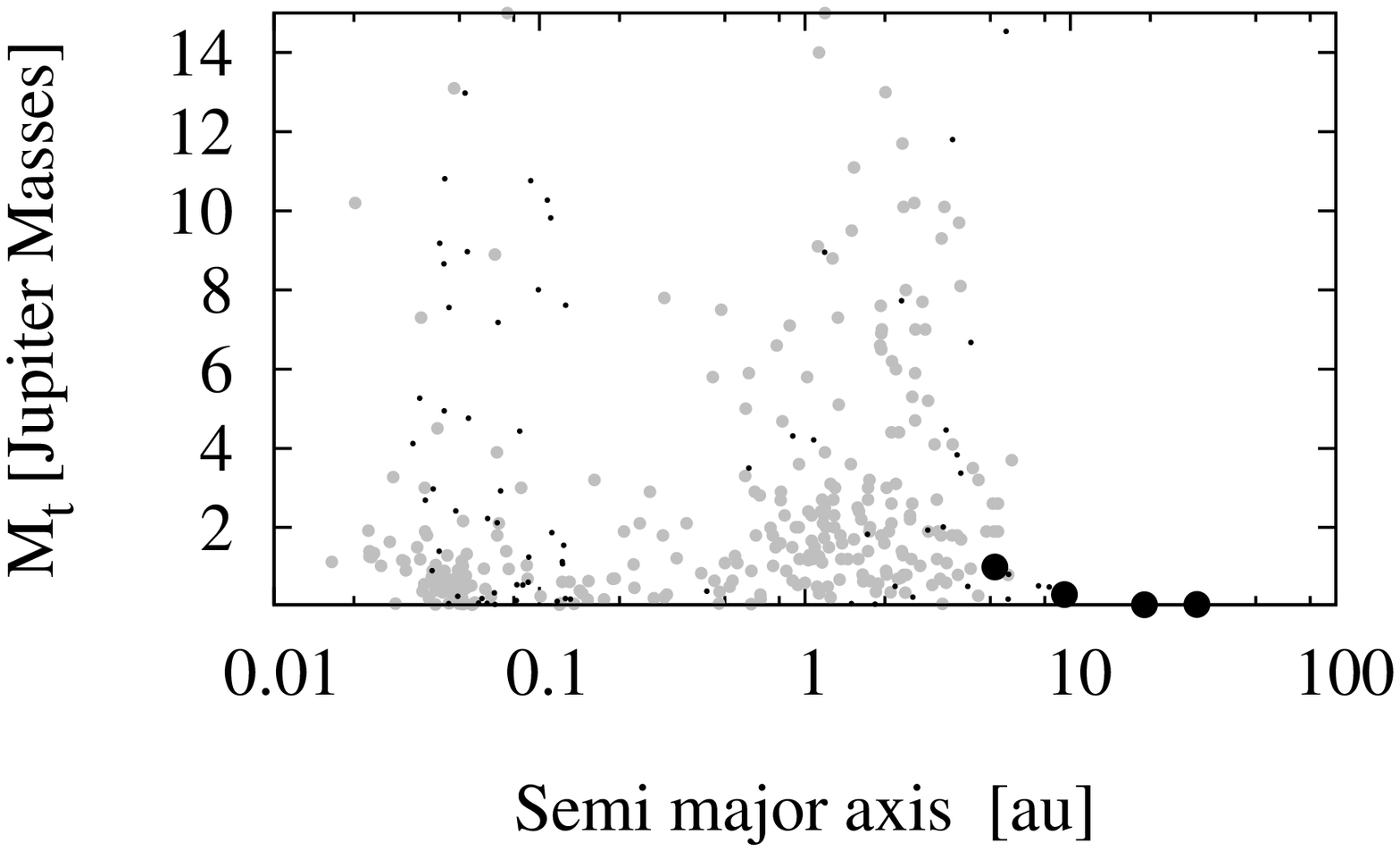}}
    \subfigure[]{\label{3comp-c1}\includegraphics[angle=270,width=.48\textwidth]{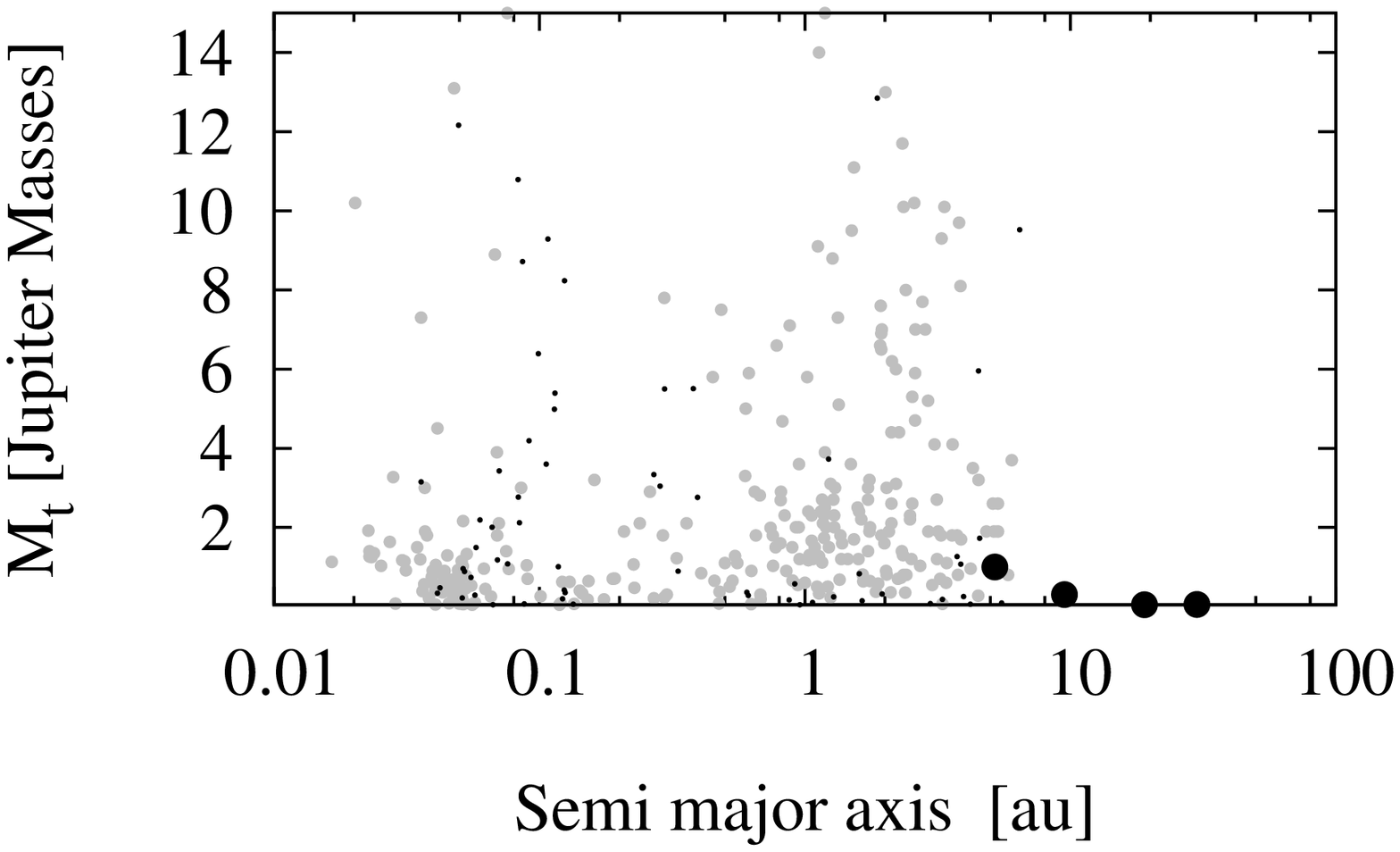}}
  \end{center}
  \caption{In the figures the mass and semi major axis distribution of our simulation results when the migration is considered and it was not delayed ($c_{migI}=1$) are shown with the small black dots, the distribution of observed exoplanets are overlapped in grey dots and also Jupiter, Saturn, Uranus and Neptune are shown with the biggest black dots. Figure \ref{1comp-c1} shows the simulation results obtained with $\gamma=1.5$, Figure \ref{2comp-c1} represents the results found with $\gamma=1$ and if $\gamma=0.5$ we obtained figure \ref{3comp-c1}.}
  \label{comp-c1}
\end{figure}

In order to get a better understanding of which are the parameters that gives us the best fit to the observations we superimposed the observations on our results found without considering the migration (\ref{comp-sinmig}) and when it is delayed, considering c = 0.01 (\ref{comp-c001}), c = 0.1 (\ref{comp-c01}) and c = 1 (\ref{comp-c1}), for the three values of $\gamma$ analyzed in this work.

In Figure \ref{comp-sinmig} we show the overlap of the observations (exoplanets are grey dots and black big dots are Jupiter, Saturn, Uranus and Neptune) with the simulation results when the effect of planetary migration is not considered. The case with $\gamma=1.5$ is shown in \ref{1comp-sinmig}, where the simulation results fit pretty well the population II of exoplanets but it can not reproduce the formation of the giant planets in our Solar System. When $\gamma=1$ there are more solids available to form giant planets farther from the star and as a result the population II of the observations can be reproduced and also the giant planets in the Solar System. Finally when $\gamma=0.5$ the simulations show that the giant planets are formed preferably about $10~au$ and although this allows the formation of the giant planets in the Solar System, does not reproduce well the observed distribution of extrasolar planets. As a conclusion, the value $\gamma=1$, which is the approximate medium value found by \citet{b2} on their observations, are the best value in order to reproduce the population II of observed exoplanets and the giant planets in the Solar System, but it fails reproducing the population I of exoplanets.  

In order to try to form the planets belonging to population I, we introduce the effects of orbital migration of planets. Figure \ref{comp-c001} shows the results for different values of $\gamma$ when the migration is considered, but the type I migration was slowed down 100 times ($c_{migI}=0.01$). Figure \ref{1comp-c001} shows the results with $\gamma=1.5$, when $\gamma=1$ we found figure \ref{2comp-c001}, and the third figure was obtained when considering $\gamma=0.5$. We can notice that population I has began to appear, but this is stronger in the first (figure \ref{1comp-c001}) and in  the second one (Figure \ref{2comp-c001}).

In the next set of figures (\ref{comp-c01}) we show the overlap of the observations (plus the giant planets in the Solar System) and our simulation results when the migration is considered and the parameter for delaying type I migration is $c_{migI}=0.1$. When $\gamma=1.5$ we observe figure \ref{1comp-c01}, when $\gamma=1$ we obtained \ref{2comp-c01} and with $\gamma=0.5$ we found the overlap shown in figure \ref{3comp-c01}. As was explained in the previous section, in these figures the same two populations noticed in the observations appeared, which are a consequence of planetary migration. Nevertheless these two populations are best differentiated in Figures \ref{1comp-c01} and \ref{2comp-c01}, which means for $\gamma=1.5$ and $\gamma=1$. 

Although these two figures show the results that are the best fit to the observations, the presence of a large population of very massive giant planets (with masses larger than $5~M_{Jupiter}$) is noticed here and it is not observed. If we calculate the percentage of observed exoplanets with masses greater than $5~M_{Jupiter}$ and located at a distance $\le 0.4~au$, we find that the 2.37\% of the observed exoplanets have these characteristics, while the same percentage calculated for the results of our simulations shows that the number of super-hot Jupiters for the case where $\gamma=1.5$ and $c_{MigI}=0.1$ is 30.86\% and the percentage is 18.58\% when $\gamma=1$ and $c_{migI}= 0.1$. We could think that these super hot-Jupiter population exists but it was not detected yet. But, if such planets exist, they should have been observed, since their observation is favored by the observational techniques, so these overpopulation must represent a limitation of our model. We will discuss this in section \ref{discusion}.

If there is no factor delaying type I migration the overlap of observations and simulation results are those shown in figure \ref{comp-c1}. The results obtained with $\gamma=1.5$ are shown in Figure \ref{1comp-c1}, those obtained with $\gamma=1$ are represented in Figure \ref{2comp-c1} and the last Figure show the results when $\gamma=0.5$. In this case we note that the migration is too fast, and most of the planets reach the inner edge of the disc, so with this value of $c_{migI}$ we can reproduce the population I of exoplanets observed but the population II is absent.    

In summary, we find that when we do not consider planetary migration we can only reproduce population II and when migration is considered but it is not delayed,  then only population I is reproduced. As a conclusion, planetary migration is a critical factor to explain the observed distribution and we need to get a better understanding of the factors that act slowing down and stopping planetary migration in order to get a better explanation of the planetary formation process.

On the other hand one of the main objectives of our work is to show the importance of considering different protoplanetary nebula profiles on giant planet formation. If we analyze which is the disc model that better explains the observations, we found that both $\gamma=1.5$ and $\gamma=1$ represent them quite well, but we choose $\gamma=1$ as the value that allows us to have an initial protoplanetary disc consistent with the observations and capable of reproduce the exoplanets observational sample and also the four giant planets of our Solar System.

\subsection{About the overpopulation of massive giant planets close to the star}\label{discusion}

As discussed in the previous section, although we obtained a good fit to the observations (Figure \ref{2comp-c01}), we found an overabundance of planets with masses greater than $5~M_{Jupiter}$ and located very close to the central star, which are not detected observationally. In order to try to determine the cause of this overabundance of giant planets in the inner edge of the disc, we explored different alternatives. On the one hand we analyzed the possibility that such planets may have lost much of its atmosphere by erosion due to stellar winds and as a result most of its primordial gaseous envelope has escaped. According to \cite{b44}, the mass loss of hot-Jupiters due to this effect is $\sim 10^{12}~g~s^{-1}$ or $ \sim 1.5 \times 10^{-2}~M_{Jup}~Gy^{-1}$, which is an important effect for objects with masses less or equal than Jupiter, but is irrelevant for super hot-Jupiters and therefore does not explain the observed mass excess.

On the other hand we also analyzed the possibility that the dissipation of gas in the nebula has a significant effect on these super hot-Jupiters. The gaseous disc dissipates between 1 and 10 million years \citep{b41,b42} due to different effects, one of them is the viscous draining on to the central star \citep{b5}, although the photoevaporation is also very important \citep{b43,b27}. In this paper we considered a simple model of gas dissipation in the disc, where the gas dissipates exponentially but independently of the distance to the central star, while in reality, the gaseous dissipation is much more complicated. \cite{b27} have shown that a low level of photoevaporative mass loss from the disc in the region between $5$ and $10~au$ can promote the eventual rapid switch off of the disc inward of this radius, this coincides with the epoch at which the accretion rate through the disc falls as a consequence of viscous draining, to a level comparable with the photoevaporative mass-loss rate. As a result the inner disc empties faster than the rest of the disc. This phenomena could stop the type II migration of a giant planet already formed. So this could be one of the possible reasons of why we do not observe very massive giant planets close to that star. In order to explore this effect we need a more rigorous model for the disc evolution, which will be incorporated in our following works.

Finally, when comparing figure \ref{2comp-c01} with other plots obtained with different values to delay the migration, there is no overpopulation when migration is slower (Figure \ref{2comp-c001}), so this effect could also be an effect of the $c_{migI}$ value considered.

When considering simultaneously the standard core accretion scenario for giant planet formation and gaseous migration for the orbital evolution, we face a serious problem. For standard disc models, type I migration are primarily inward, and their timescale is at least 1 to 2 orders of magnitude shorter than the disc lifetime ($10^6-10^7$ yr), which means the cores would be accreted by the central star before they could build up any substantial gaseous envelope. Therefore, in order to avoid being swallowed by their central stars, the planet's migration must be significantly slowed down or stopped somehow, to this end we used the factor $c_{migI}$, as was explained on section \ref{migracion}. But the question is, how much should we slow down planetary migration? 

A number of scenarios have been proposed to solve the non stop migration problem.  \citet{b26} explore how the uncertainties in the structure of protoplanetary discs may affect type I migration rates and found that it can be significantly slowed down at opacity transitions in the disc, \citet{b30} show that as the surface density and temperature fall the planet orbital migration and disc depletion timescales eventually become comparable, in \citet{b31} it was shown through three-dimensional, radiation-hydrodynamic simulations that planets could suddenly move outward as well as inward, depending on the local opacity, \citet{b32} proposed that the low-viscosity regions in protostellar discs (dead zones) can significantly slow down planet orbital motion, \citet{b33} and \citet{b34} shown that including magnetic fields may slow down or even stop type I migration, \citet{b35} showed that surface density jumps in the disc can trap low-mass protoplanets reducing the type I migration rate to the disc's accretion rate. We also add that irradiation should lead to a hole formed by photo evaporating wind \citep{b27,b28,b29}, which quickly depletes the disc, which possibly bring migration to an even earlier halt.

All these works show that type I migration rate strongly depends on the exact disc conditions in the vicinity of the embedded embryo. Therefore we must be cautious about what is the $c_{migI} $ value assumed. The factor that slows down migration should be a parameter that depends on the disc conditions in the proximity of the embryo and as such it is expected to be not a unique value for all discs but a distribution of possible values depending on the initial disc characteristics. We also propose that this could be a possible solution to the problem of the overabundance observed, since the number of massive giant planets observed in population I, is strongly dependent on the value of $c_{migI}$ considered, in other words, on the rate of type I migration assumed. So in order to have a more appropriate model, we should consider a distribution of values $c_{migI}$, taking into account the factors that inhibit the migration depends on the disc where the planets are embedded.

\section{Summary and Conclusions}\label{conclusion}

In the present work we have developed a semi-analytical model for computing planetary systems formation which is based on the core instability model for the gas accretion of the embryos and the oligarchic growth regime for the accretion of the solid cores. The gas accretion model considered is a new approximation that we obtained introducing an analytic approximation to the numerical results found by \citep{b19}. 

The disc surface density profiles plays a fundamental role in the formation of giant planets and therefore, in the formation of the Planetary System, so in order to get a clear understanding of which are the initial conditions that allowed the formation of the observed exoplanets and the giant planets in the Solar System we explore different models for the initial protoplanetary nebula structure. We assume that the gas and solid surface density are characterised by a power-law in the inner part of the disc and an exponential decay in the outer parts, which has great advantages, one of which is that we are not forced to arbitrarily cut the disc in order to obtained a finite disc mass. Both discs change with time: the solid disc is locally depleted due to the accretion of the embryos and the gas disc is globally depleted with a time-scale between $10^6$ and $10^7$ years. 

In our model we assume that the embryos have an orbital motion due to their interaction with the host star and consider two regimes of planetary migration: type II for the larger embryos and type I for the smaller ones, where the rate is delayed a factor $c_{migI}$ in order to represent the factors that act slowing down or even stopping planetary migration. 

With this model we consider different initial conditions to generate a variety of planetary systems and analyse the giant planets statistically. We assumed different initial discs profiles, a large range of discs masses and sizes according to the last results of observations in protoplanetary discs \citep{b2,b3}, different stars in a range between $0.7$ and $1.4 M_{\odot}$, and also different values of the parameter which acts delaying type I migration. We explore the effects in the formation of assuming different discs and found that the observed population of exoplanets, including Jupiter, Saturn, Uranus and Neptune, is well represented when considering planetary migration and a surface density with a power-law in the inner part characterised by an exponent of $\gamma=1$, which represents a softer profile when compared with the case most similar to the MMSN model case. Nevertheless we would like to note that despite having chosen $\gamma=1$ as the best value, in reality we should have considered a distribution of values of $\gamma$ in order to take into account the different discs where the planetary formation can occur. But, since the media value of the observations is $\gamma \sim 1$ (Andrews et al., 2009), the fact of not considering a distribution does not represent a big problem in the model considered.

On the other hand if we explore the effects of considering different values for delaying type I migration, we found that the best match to the observations is obtained when we delayed it some value between $10$ and $100$ times, but as noticed in section \ref{discusion}, this parameter depends strongly on the disc characteristics and therefore should be taken from a distribution of values and not be a unique value as has been considered so far in the synthetic population models. So we should have a better understanding of the factors that acts slowing down and stopping planetary migration in order to get a better explanation of the planetary formation process.

\end{document}